\documentclass[fleqn,usenatbib]{mnras}

\usepackage{newtxtext,newtxmath}

\usepackage[T1]{fontenc}

\DeclareRobustCommand{\VAN}[3]{#2}
\let\VANthebibliography\thebibliography
\def\thebibliography{\DeclareRobustCommand{\VAN}[3]{##3}\VANthebibliography}

\usepackage{graphicx}	
\usepackage{amsmath}	

\usepackage{amssymb}	
\usepackage{verbatim}
\usepackage{multirow}
\usepackage{threeparttable}
\usepackage{relsize}
\usepackage{color}

\usepackage{caption}
\usepackage{subcaption}

\usepackage[splitrule]{footmisc} 
\newcommand{\angstrom}{\textup{\AA}}
\newcommand{\Msun}{\mathrm{M_\odot}}

\newcommand{\km}{\mathrm{km}}

\newcommand{\s}{\mathrm{s}}
\newcommand{\yr}{\mathrm{yr}}

\newcommand{\fesclya}{f_{\rm esc, Ly\alpha}}
\newcommand{\fesclyc}{f_{\rm esc, LyC}}

\newcommand{\vred}{v_{\rm red}}

\newcommand{\sigred}{\sigma_{\rm red}}
\newcommand{\NHI}{N_{\rm H~\textsc{i}}}

\newcommand{\vvir}{v_{\rm vir}}
\newcommand{\xhi}{x_{\rm H~\textsc{i}}}
\newcommand{\LLya}{L_{\rm Ly\alpha}}
\newcommand{\lya}{Ly$\alpha$\ }

\defcitealias{Witten24}{W24}
\defcitealias{Yuan_Yuxuan24}{Y24}

\usepackage[normalem]{ulem}
\definecolor{darkgreen}{rgb}{0.13, 0.55, 0.13}
\definecolor{orange}{rgb}{0.8, 0.15, 0.13}

\newcommand{\aref}[1]{\hyperref[#1]{Appendix~\autoref{#1}}}

\usepackage{scalerel,tikz}
\usetikzlibrary{svg.path}
\definecolor{orcidlogocol}{HTML}{A6CE39}
\tikzset{orcidlogo/.pic={
 \fill[orcidlogocol] svg{M256,128c0,70.7-57.3,128-128,128C57.3,256,0,198.7,0,128C0,57.3,57.3,0,128,0C198.7,0,256,57.3,256,128z};
 \fill[white] svg{M86.3,186.2H70.9V79.1h15.4v48.4V186.2z}
 svg{M108.9,79.1h41.6c39.6,0,57,28.3,57,53.6c0,27.5-21.5,53.6-56.8,53.6h-41.8V79.1z M124.3,172.4h24.5c34.9,0,42.9-26.5,42.9-39.7c0-21.5-13.7-39.7-43.7-39.7h-23.7V172.4z}
 svg{M88.7,56.8c0,5.5-4.5,10.1-10.1,10.1c-5.6,0-10.1-4.6-10.1-10.1c0-5.6,4.5-10.1,10.1-10.1C84.2,46.7,88.7,51.3,88.7,56.8z};
}}
\newcommand\orcidicon[1]{\href{https://orcid.org/#1}{\mbox{\scalerel*{
\begin{tikzpicture}[yscale=-1,transform shape]
\pic{orcidlogo};
\end{tikzpicture}
}{|}}}}

\title[Visibility of LAEs]{Extended red wings and the visibility of reionization-epoch Lyman-$\alpha$ emitters}

\author[Yuan et al.]{
Yuxuan Yuan$^{1}$~\orcidicon{0000-0001-6816-0682}\thanks{E-mail: yy503@cam.ac.uk (YY)}, Sergio Martin-Alvarez$^{2}$~\orcidicon{0000-0002-4059-9850}, Martin G. Haehnelt$^{1}$, Thibault Garel$^{3}$, 
\newauthor Laura Keating$^{4}$, Joris Witstok$^{5,6}$~\orcidicon{0000-0002-7595-121X} and Debora Sijacki$^{1}$
\\
$^{1}$ Institute of Astronomy and Kavli Institute for Cosmology, University of Cambridge, Madingley Road, Cambridge CB3 0HA, UK\\
$^{2}$ Kavli Institute for Particle Astrophysics \& Cosmology (KIPAC), Stanford University, Stanford, CA 94305, USA\\
$^{3}$ Observatoire de Genève, Université de Genève, Chemin Pegasi 51, 1290 Versoix, Switzerland\\
$^{4}$ Institute for Astronomy, University of Edinburgh, Blackford Hill, Edinburgh, EH9 3HJ, UK\\
$^{5}$ Cosmic Dawn Center (DAWN)\\
$^{6}$ Niels Bohr Institute, University of Copenhagen, Jagtvej 128, 2200 København N, Denmark
}

\date{MNRAS, submitted}

\pubyear{2024}

\begin{document}
\label{firstpage}
\pagerange{\pageref{firstpage}--\pageref{lastpage}}
\maketitle

\begin{abstract}
The visibility of the Lyman-$\alpha$ (Ly$\alpha$) emission from reionization-epoch galaxies depends sensitively on the extent of the intrinsic \lya emission redwards of 1215.67~\AA. The prominent red peak resulting from resonant radiative transfer in the interstellar medium is often modelled as a single Gaussian.  We use the \textsc{Azahar} simulation suite of a massive-reionization epoch galaxy to show that a significantly larger fraction of the \lya emission extends to $400$-$800$~km~s$^{-1}$, and thus significantly further to the red than predicted by a Gaussian line profile. A cycle of frequent galaxy mergers strongly modulates the \lya luminosity, the red peak velocity and its extended red wing emerging from the galaxy, which all also strongly vary with viewing angle. The \lya emission also depends sensitively on the implemented feedback, dust and star formation physics. Our simulations including cosmic rays reproduce the observed spectral properties of reionization epoch \lya emitters (LAEs) well if we assume that the \lya emission is affected by very little dust. The visibility of LAEs can be strongly underestimated if the extended red wings of the intrinsic \lya emission are not accounted for. We discuss implications for using the visibility of LAEs to constrain the evolution of the volume-averaged neutral fraction during reionization. 
\end{abstract}



\section{Introduction}

The Universe transits from neutral to ionized hydrogen during the epoch of reionization (EoR; see e.g. \citet{Robertson22} for a recent review). The reionization history of the intergalactic medium (IGM) is also of crucial importance to galaxy formation and evolution. The electron scattering optical depth measured against the cosmic microwave background (CMB; \citealt{Planck_Col16_reion, Planck_Col20}) tells us that the reionization process has already progressed significantly at $z\sim 7$ while the \lya forest in QSO absorption spectra \citep{Kulkarni19_late_reion, Keating20_long_troughs, Bosman22} suggests that reionization does not complete until $z\sim 5.3$.

The Ly$\alpha$ line is a very luminous emission line and has been observed up to $z\sim 13$ \citep{Ouchi20, Witstok24_lae_z13}. Neutral atomic hydrogen (H~\textsc{i}) has a rather large scattering cross section at \lya wavelengths and therefore the visibility of Ly$\alpha$ emitters (LAEs) changes drastically during the transition from a neutral to an ionized IGM. Detections of LAEs are common at $z\sim 6$ \citep{Stark11, De_Barros17}, but decline at $z\gtrsim 7$ \citep{Stark10, Treu13, Hoag19, Jung20} due to the Ly$\alpha$ damping wing of the intervening increasingly neutral IGM \citep{Miralda-Escude98, Laursen11, Mesinger15, Umeda24}. 

The advent of JWST has allowed a large population of high-$z$ LAE to be observed, including the iconic GN-z11 \citep{Bunker23}. \citet{Tang_Mengtao23} and \citet{Saxena24} found that high-$z$ ($z\gtrsim 7$) LAEs show lower escape fraction, $\fesclya$, and larger \lya velocity offsets than low-$z$ samples, indicating increasingly smaller sizes of the surrounding ionized bubbles. \citet{Tang_Mengtao24_lya_oiii} showed that high-$z$ LAEs typically have Ly$\alpha$ velocity offsets of about $230$~km~s$^{-1}$, larger than those of $z\sim 2-3$ LAE, and reflecting the effect of IGM attenuation. \citet{Chen_Z24} and \citet{Saxena23} on the other hand discovered high-$z$ LAEs with near-unity IGM transmission, as well as LAEs with $\fesclya \sim 1$. They also report a faint LAE with an extremely high equivalent width within an ionized bubble at $z = 7.3$. \citet{Witstok25_pri_lae} discovered three high-redshift ($z\gtrsim 8$) low-mass LAEs with low Ly$\alpha$ velocity offsets and varying levels of EWs. These galaxies show signatures of recent, vigorous starbursts and are surrounded by ionized bubbles of various inferred sizes (the one with the largest EW is located within a bubble with an inferred size of $\sim$ 3~pMpc). The observational evidence regarding the decline of the visibility of LAEs at high redshift is thus complex.

Ly$\alpha$ emission escaping a galaxy will suffer attenuation by the circumgalactic medium (CGM) as well as the  IGM. \citet{Sadoun17} found that the CGM out to 10~$r_{\rm vir}$ can also strongly attenuate the Ly$\alpha$ emission near the line centre. The effect of IGM attenuation has been studied extensively with analytic models \citep{Miralda-Escude98, Dijkstra07, Inoue14, Mason&Gronke20} as well as semi-analytic models such as the \textsc{Dragons} model \citep{Qin22} and \textsc{Galform} \citep{Gurung-Lopez19_lya_cos, Gurung-Lopez20}. More self-consistent studies use cosmological simulations to relate the large-scale distribution of the neutral IGM and the corresponding attenuation \citep{Zheng10, Laursen11}. Recently, more realistic modelling of reionization and the IGM attenuation calibrated to \lya forest data has become possible, such as that based on the \textsc{Sherwood} simulation \citep{Weinberger_L18, Weinberger_L19}, the \textsc{Sherwood-Relics} simulations \citep{Keating24_ori_shape_scat_dw, Keating24_jwst_dw_cos}, as well as the \textsc{Sphinx} \citep{Garel21}, \textsc{CoDaII} \citep{Gronke21, Park_H21b}, and \textsc{Thesan} \citep{A_Smith22a} simulations. 

The visibility of LAEs is expected to trace the reionization history of the Universe \citep{Dijkstra&Wyithe10}, and \citet{Gurung-Lopez20, Garel21, A_Smith22a} showed how IGM transmission can be the dominant driver of the apparent redshift evolution of LAEs within a realistic cosmological radiative transfer simulation. High-$z$ LAEs are thus used to infer the evolution of the neutral fraction of the IGM (\citealt{Stark10, Stark11, Pentericci14, Mason18_baye_igm}, also see \citealt{Tang_Mengtao24_lya_z6p5_13, Jones25} for recent studies based on JWST data). Numerous recent works have built hybrid Bayesian frameworks to connect observed high-$z$ LAEs with predictions from numerical simulations of the IGM, aiming to gain a quantitative description for the neutral IGM surrounding the LAE population \citep{Mason18_baye_igm}. For example, \citet{Hayes23c} proposed a Bayesian framework to calculate the bubble size from observed Ly$\alpha$ spectra. These works, however, probably underestimate the degeneracies between \lya radiative transfer effects due to scattering on ISM and CGM/IGM scales \citep{Sadoun17, Hassan&Gronke21}.

The attenuation by the IGM/CGM and thus the visibility of LAEs during the epoch of reionization depends strongly on how far the Ly$\alpha$ emission extends to the red after scattering by the ISM. \citet{Mason&Gronke20} used the minimum observed Ly$\alpha$ velocity shift to put a lower limit on the bubble size and an upper limit on the neutral hydrogen fraction within ionized bubbles. \citet{Weinberger_L19} assumed Ly$\alpha$ velocity offsets of $\sim 250$~km~s$^{-1}$ when fitting the observed Ly$\alpha$ luminosity function (LF) at $z\sim 7$. 

Both theoretical and observational studies have shown that galaxy mergers trigger intense star formation and feedback. This means galaxy mergers are a possible mechanism that boosts production and escape of \lya and ionizing photons \citep{Rauch11}, and thus aid reionization \citep{Cooke10, Jimenez-Andrade23, Solimano22, Jung24}. The observed examples include Himiko \citep{Ouchi09} and CR7 \citep{Sobral15}. \citet{Qin22} and \citet{Witten24} found that bright  high-$z$ galaxies with a companion will have higher Ly$\alpha$ visibility. This phenomenon is also observed in the local Green Pea galaxy GP J0213+0056 \citep{Purkayastha22}. However, note that several works report no correlation between merger signature and \lya/LyC escape, including the local Ly$\alpha$ Reference Sample (LARS; \citealt{Micheva18}), an observational sample at medium redshift of $z\sim 2.2$ \citep{Shibuya14_lae_demo}, and a high-$z$ sample observed with JWST \citep{Mascia25}.

The paper builds upon the methods and analysis pipeline of \citet[hereafter \citetalias{Yuan_Yuxuan24}]{Yuan_Yuxuan24} and extends the analysis in \citet[hereafter \citetalias{Witten24}]{Witten24}. Up to now, few theoretical works have focused on Ly$\alpha$ emission from merging galaxies and its impact on inferring IGM properties. In this paper, we aim to address the following questions: (1) What is the impact of the merger sequence, LOS statistics, and different simulated physics on the \lya observables from high-$z$ LAE? (2) What are the connections between our simulated merger and the recently observed high-$z$ LAE populations? (3) Are there any caveats to the common methods of LAE-based inference of the IGM neutral fraction?

The structure of the paper is as follows: in \autoref{sec:methods} we describe the simulations and our post-processing technique. Next, we explore the variations of the merger sequence due to different simulated physical processes and compare them to JWST observations in \autoref{sec:res}. We discuss several caveats regarding our results in \autoref{sec:disc} and give our conclusions in \autoref{sec:con}.

\section{Methods}
\label{sec:methods}

In this section, we give an overview of the numerical simulations and post-processing techniques that we use. We first describe the \textsc{Azahar} simulation suite in \autoref{ssec:sim}. We summarize our \textsc{Rascas} post-processing technique in \autoref{ssec:emi}. We describe our treatment of IGM attenuation and JWST observational effects in \autoref{ssec:igm_attenuation}. We lastly define several diagnostic quantities frequently used in observations in \autoref{ssec:diag_quan}.

\subsection{The \textsc{Azahar} simulation suite}
\label{ssec:sim}

\begin{table}
  \centering
    \begin{tabular}{ccccc}
    \hline\hline
    Simulation name & RT & SN & Magnetic field & CR \\ \hline
    RT & \checkmark & Fiducial & - & - \\
    RTiMHD & \checkmark & Fiducial & Astrophysical & - \\
    RTnsCRiMHD & \checkmark & Fiducial & Astrophysical & no streaming \\
    RTCRiMHD & \checkmark & Fiducial & Astrophysical & \checkmark \\
    \hline\hline
    \end{tabular}
  \caption{Summary of \textsc{Azahar} simulation runs used in this paper.}
  \label{tab:sim_tab}
\end{table}

In this work, we use cosmological zoom-in simulations from the \textsc{Azahar} suite (Martin-Alvarez, in preparation). The simulations have been performed with the cosmological hydrodynamical radiative transfer code \textsc{Ramses-rt} \citep{Teyssier02, Rosdahl13}. \textsc{Azahar} simulates a massive disc galaxy, with a halo mass of $\sim 2\times 10^{12}\,\Msun$ at $z=1$. The size of the zoom-in region is $\sim$~8~cMpc across. Gas cells refine when their enclosed mass is larger than 8 times the dark matter particle mass or when their sizes are larger than 4 times the local Jeans length. The highest resolution level in the simulation corresponds to a spatial resolution of $20$~pc. In the zoom-in region, dark matter particles have a mass $m_{\rm DM} = 4.5 \times 10^5\,\Msun$ and the mass resolution of star particles is $m_* \approx 4\times 10^4\,\Msun$.

The physical processes included in the simulations are similar to those in its pathfinder simulation suite \textsc{Pandora} \citep{Martin-Alvarez23} and we briefly summarize them here. Radiative cooling includes the primordial and metal cooling of hot gas ($\gtrsim 10^4$~K) using tables from \textsc{Cloudy} \citep{Ferland98}, as well as the metal cooling of cold gas following \citet{Rosen&Bregman95}. The simulations adopt the \citet{Haardt&Madau96} ultra-violet background (UVB) model and assume ideal monatomic gas with an adiabatic index of $\gamma = 5/3$. The simulations further adopt a Schmidt law for star formation \citep{Schmidt59}, $\Dot{\rho}_{\rm star} = \epsilon_{\rm ff} \rho_{\rm g}/t_{\rm ff}$, where $\epsilon_{\rm ff}$ is the star formation efficiency and is calculated assuming  the magneto-thermo-turbulent (MTT) star formation model described in \citet{Kimm17} and \citet{Martin-Alvarez20}. The mechanical supernovae (SN) feedback prescription \citep{Kimm&Cen14, Kimm15} attempts to recover the correct terminal momentum of the snowplough phase of the supernovae remnants (SNR) even when the Sedov-Taylor phase is not resolved. The radiative transfer modelling follows the implementation in \textsc{Ramses-rt} \citep{Rosdahl13, Rosdahl&Teyssier15}. It accounts for photon injection, photo-ionization, photo-heating and radiation pressure. The M1 closure scheme is used to solve the coupled radiation-hydrodynamical equations. Three frequency bins corresponding to the ionization of H~\textsc{i} ($13.6\,{\rm eV} \lesssim \epsilon_{\rm photon} \lesssim 24.59\,{\rm eV}$), He~\textsc{i} ($24.59\,{\rm eV} \lesssim \epsilon_{\rm photon} \lesssim 54.42\,{\rm eV}$), and He~\textsc{ii} ($54.42\,{\rm eV} \lesssim \epsilon_{\rm photon}$) are used. Because of the explicit treatment of advection, a reduced speed of light $\tilde{c} = 0.01c$ is adopted to avoid the otherwise extremely small time steps\footnote{This approximation is valid when the reduced light speed $\tilde{c}$ is much higher than the propagation speed of the ionization front \citep{Rosdahl13, Rosdahl&Teyssier15, Katz17}. According to the calculation in Section 4.3 of \citet{Rosdahl13}, the reduction factor should be $\sim 10^{-2}$ for ISM conditions.}. We subcycle the radiation solver up to a maximum of 500 steps over one step of the hydrodynamical solver. The MHD solver uses the constrained transport scheme (CT; \citealt{Fromang06, Teyssier06}), which satisfies the divergence condition $\nabla \cdot \Vec{B} = 0$ to numerical precision. For the magnetic seeding model, small-scale circular loops of the magnetic field are injected by SNe with magnetic energy corresponding to  1 per cent of the energy of SNe, $E_{\text {ini,mag}} = 0.01 E_{\mathrm{SN}} = 10^{49} \mathrm{erg}$ \citep{Martin-Alvarez21}, to reproduce the magnetisation of SN remnants at small scales \citep{Parizot06}, and yielding realistic galactic magnetisations \citep{Martin-Alvarez24a}. The cosmic ray (CR) module \citep{Dubois16, Dubois19} models CR anisotropic diffusion and streaming and incorporates hadronic and Coulomb cosmic ray cooling \citep{Guo&Oh08}. We adopt a constant diffusion coefficient (along magnetic field lines) of $\kappa_{\parallel} = 3\times 10^{28} \mathrm{cm}^2 \mathrm{s}^{-1}$, motivated by $\gamma$-ray observations of cosmic ray hadronic losses \citep{Ackermann12, Salem16, Pfrommer17_gamma_ray} and their Milky Way isotropic coefficient \citep{Trotta11, Cummings16}. Each SN injects a fraction $f_{\rm CR}=0.1$ of its energy as  CR energy, $E_{\rm CR} = f_{\rm CR} E_{\rm SN} = 10^{50} \mathrm{erg}$ \citep{Morlino&Caprioli12, Helder13}.

The \textsc{Azahar} suite contains a set of simulations with different complexity of the physical processes included. The name of each simulation specifies its included physical processes and configuration. Radiative transfer is represented by ``RT'' and the magnetic field configuration is represented by ``iMHD''. The CR configurations consist of a full physics run ``CR'' and a run with CR streaming switched off, ``nsCR''. Here, we focus on four simulation runs RT, RTiMHD, RTnsCRiMHD, and RTCRiMHD from \textsc{Azahar}. A summary of the four simulations is given in \autoref{tab:sim_tab}.  RTCRiMHD should be considered the most complete simulation in this study (see Martin-Alvarez, in preparation, for more details).

We track the movement and merging events of the galaxies in the simulations with an updated version of the tracker algorithm originally described in \citetalias{Witten24} and \citet{Sanati24}, which we briefly describe here. We identify three progenitor galaxies with the {\sc Halomaker} software \citep{Tweed09} and follow them through time by tracking their innermost 500 stellar particles, or $0.4 M_* / m_*$ (with $M_*$ the mass of each galaxy at a given time), whichever is lower. This algorithm determines the centres with a shrinking-spheres algorithm \citep{Power03} applied to the stellar component. For each galaxy, our tracker further determines the physical properties of individual galaxies within the galactic region, defined by the radius $r_{\rm gal} < 0.2 r_{\rm vir}$, where $r_{\rm vir}$ is the virial radius as measured by {\sc Halomaker}. During the merging stages, we separate in our measurements each merging galaxy by employing $r < {\rm min}(r_{\rm gal}, 0.45D_{ij})$, where $D_{ij}$ is the distance between the $i$ and $j$ progenitors, with a minimum distance of $1.5$~kpc.
We differentiate between: 
\begin{itemize}
    \item an isolated phase, where all three galaxies are well separated,
    \item a merger phase, where some of the galaxies  are merging with others,
    \item and a disc phase, where all three galaxies have merged into one single massive disc galaxy.
\end{itemize}

\subsection{Ly\texorpdfstring{$\alpha$}{alpha} and LyC emission}
\label{ssec:emi}

In order to model the emission and propagation of Ly$\alpha$ photons, we post-process our simulation using the publicly available, massively parallel, Monte Carlo radiative transfer code \textsc{Rascas} \citep{Michel-Dansac20}. \textsc{Rascas} first extracts from the simulations the essential information about the radiation sources and the medium where the radiation travels. Then it runs the radiative transfer simulations and generates observables. 

We extract a sphere as the computational domain for Ly$\alpha$ radiative transfer. The centre of the sphere is the merger centre (defined as the non-weighted mean of the centres of the three galaxies) in the isolated/merger phase and the centre of the merged galaxy in the disc phase. The radius of the sphere $R$ is $2\,\mathrm{max}(r_i) + 10$~kpc for the isolated/merger phase, where $r_i$ is the distance of the galaxy centre to the merger centre, and it is $10$~kpc in the disc phase. We model three sources of Ly$\alpha$ emission - recombination and collisional excitation from gas, as well as the stellar continuum near the Ly$\alpha$ wavelength. \textsc{Rascas} uses a Monte-Carlo approach to sample photon packets from the full photon distribution. In each gas cell, the PDF of the gas-frame frequency of \lya photon packets is a Gaussian with the width set by the thermal broadening $\Delta v_{\rm D}$. \lya photons change their outgoing direction relative to the incoming direction according to a phase function.  The phase function depends on the frequency of the photon in the scattering frame. \textsc{Rascas} adopts  phase functions  as described in  \citet{Hamilton40} and \citet{Dijkstra&Loeb08} for the core scattering of \lya photons and adopts Rayleigh scattering for \lya photons in the line wings. To reduce unnecessary computational overhead in high H~\textsc{i} opacity environments, \textsc{Rascas} adopts the core-skipping algorithm of \citet{Smith_Aaron15} (their Section 3.2.4), building upon the works of \citet{Ahn02, Tasitsiomi06_lya_rt_cos_sim, Tasitsiomi06_res_line_mesh, Dijkstra06, Laursen09_lya_dust} to shift the photons to the line wing without local scattering in space. This algorithm selects scattering atoms with velocity (perpendicular to photon direction; $u_\perp$) beyond a critical value $x_{\rm crit}$, which is determined locally based on the optical depth (see Equation~(35) in \citealt{Smith_Aaron15}). The choices of numerical parameters in the algorithm have been tested for convergence in \citet{Smith_Aaron15} (see their Appendix Section A2). \textsc{Rascas} also includes the recoil effect and the transition due to deuterium with an abundance of $\mathrm{D/H} = 3 \times 10^{-5}$.

We adopt the ‘Small Magellanic Cloud’ (SMC) model for the \lya absorption by dust following \citet{Laursen09_lya_dust}. The dust density is calculated as $n_{\text {dust }} = \frac{Z}{Z_0} \left(n_{\rm H~\textsc{i}}+f_{\text{ion }} n_{\rm H~\textsc{ii}}\right)$, where $f_{\rm ion} = 0.01$ and $Z_0 = 0.005$. Dust can either absorb or scatter \lya photons. The dust albedo, $a_{\rm dust}=0.32$ \citep{Li&Draine01}, sets the probability of Ly$\alpha$ scattering. The scattering angle is given by the phase function of \citet{Henyey&Greenstein41} and the asymmetry parameter is set to $g = 0.73$ \citep{Li&Draine01}. 

We generate synthetic position–position–velocity (PPV) data cubes along arbitrary lines-of-sight (LOS) with the ``peeling'' algorithm \citep{Yusef-Zadeh84, Wood&Reynold99}. We collect photons within an aperture size $l$ equal to $R$ defined above. For the fiducial RTCRiMHD simulation, we generate synthetic Ly$\alpha$ observations along 108 LOS for all snapshots from $z \sim 9 \to 6$. For the other simulations (RT, RTiMHD and RTnsCRiMHD), we generate Ly$\alpha$ observables along 12 LOS because of the high computational cost. 

We self-consistently calculate the escape fraction of LyC photons along arbitrary directions with \textsc{Rascas}. Ionizing radiation is absorbed by neutral hydrogen, neutral helium \citep{Yan01}, and singly ionized helium \citep{Osterbrock&Ferland06}, and is attenuated by dust \citep{Weingartner&Draine01}. Note that \textsc{Rascas} does not take into account secondary ionization nor the re-emission of LyC photons. We also note that the same \citet{Laursen09_lya_dust} dust model is used for LyC photons.

\subsection{IGM attenuation modelling and synthetic observations}
\label{ssec:igm_attenuation}

Due to the ionisation fronts quickly expanding beyond the zoom-in sub-volume of the \textsc{Azahar} computational domain, we cannot self-consistently model the reionization process on large scales and the size of the ionized bubble surrounding our galaxies. For this reason, we use predicted IGM damping curves from a large-scale cosmological reionization simulation by \citet{Keating24_jwst_dw_cos}. The IGM damping wings have been extracted from simulations from the \textsc{Sherwood-Relics} suite of simulations \citep{Puchwein23}. We choose the \textsc{Sherwood-Relics} simulation with a late reionization history for illustrative purposes, but there is still considerable uncertainty in the reionization history as recently discussed in \citet{Asthana24_reion_aton_he}, \citet{Cain25} and references therein. We then use the reionization history of \citet{Keating24_jwst_dw_cos} to investigate the effect of the shape and extent of the red wing on the IGM attenuation of reionization-epoch galaxies, but note that our results can be easily adapted to other reionization histories.

\citet{Keating24_jwst_dw_cos} provide the median damping wings with their associated uncertainties for objects located in massive haloes. We assume that our galaxies reside in a similar environment as those haloes\footnote{In the \textsc{Sherwood-Relics} simulations, the mass range of the haloes at $z=7.4$ is about $5.4\times 10^{10} - 2.3\times10^{11}\,\Msun/h$, similar to the mass of our simulated haloes (c.f. \autoref{fig:gas_time_evo_diff_phys}).}. 
The Ly$\alpha$ radiative transfer through both ISM/CGM and IGM shows significant variations between different LOS. For simplicity, we assume these two effects are independent, i.e. the structures and morphologies of galaxies are uncorrelated with those of the large-scale ionized bubbles. The IGM damping wings of \citet{Keating24_jwst_dw_cos} are shown in \aref{asec:model_dw_inst} in \autoref{fig:igm_dw}. In  \aref{asec:model_dw_inst} we also describe in more detail our modelling of instrumental effects. In the remainder of the paper, we consider three levels of IGM transmission, the median transmission and $1\sigma$ lower (low transmission) and higher (high transmission). We have convolved the Ly$\alpha$ spectra after intrinsic scattering by the ISM/CGM with the damping wings corresponding to these three levels of attenuation to obtain IGM-attenuated Ly$\alpha$ spectra.

We further note that the peculiar velocities of the galaxies will cause a horizontal shift of the damping wing (see Figure 13 in \citealt{Gurung-Lopez20}).  The peculiar velocities of the dark matter haloes are accounted for in the predictions by \citet{Keating24_jwst_dw_cos}. We have checked that the peculiar velocities of the haloes in our simulations are usually small $\sim 10-40$~km~s$^{-1}$ and that this will have a minor effect. For this reason, we neglect these additional peculiar velocities.

\subsection{Diagnostic quantities of the Ly\texorpdfstring{$\alpha$}{alpha} spectra}
\label{ssec:diag_quan}

\begin{figure*}
    \centering
    \begin{subfigure}[b]{.82\textwidth}
     \centering
     \includegraphics[width=\textwidth]{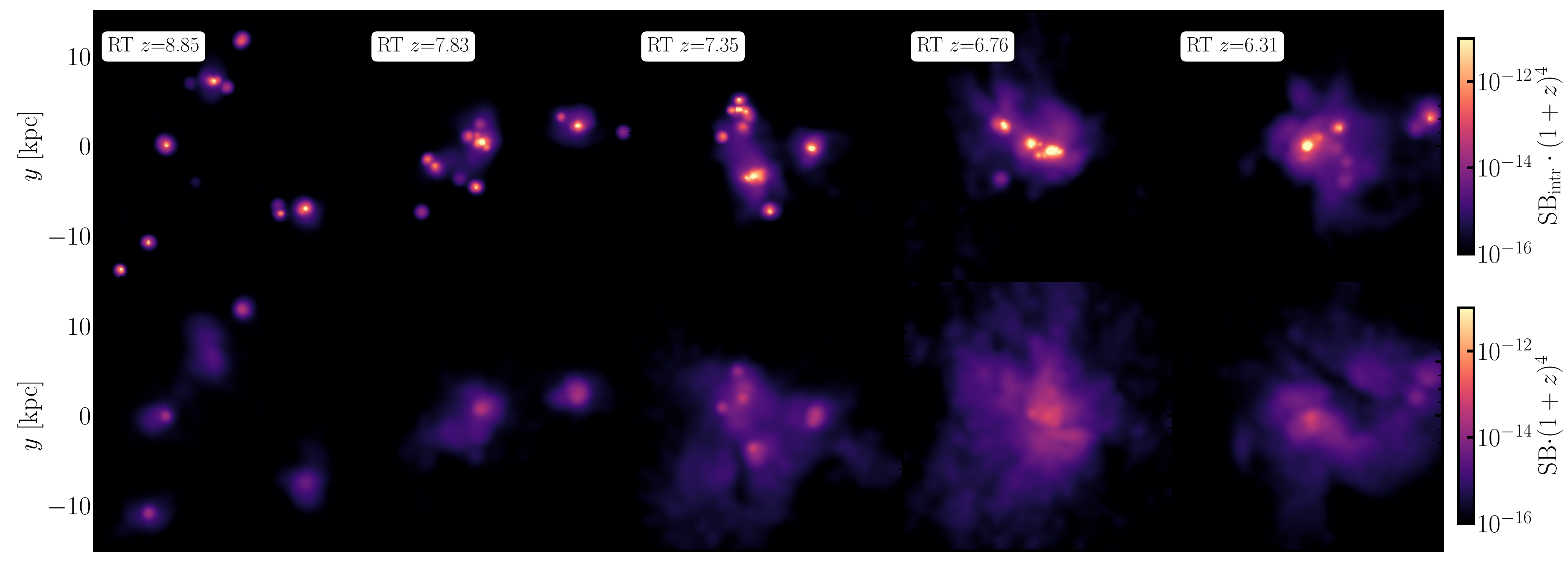}
    \end{subfigure} \\
    \begin{subfigure}[b]{.82\textwidth}
     \centering
     \includegraphics[width=\textwidth]{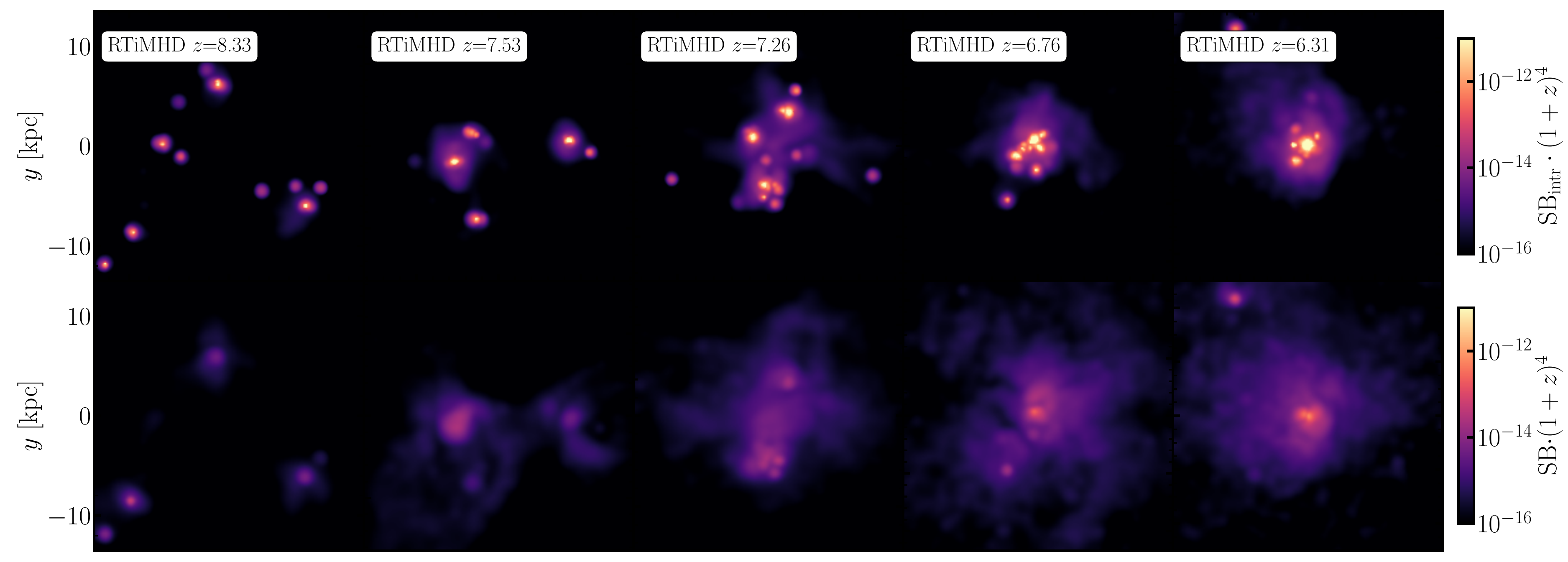}
    \end{subfigure} \\
    \begin{subfigure}[b]{.82\textwidth}
     \centering
     \includegraphics[width=\textwidth]{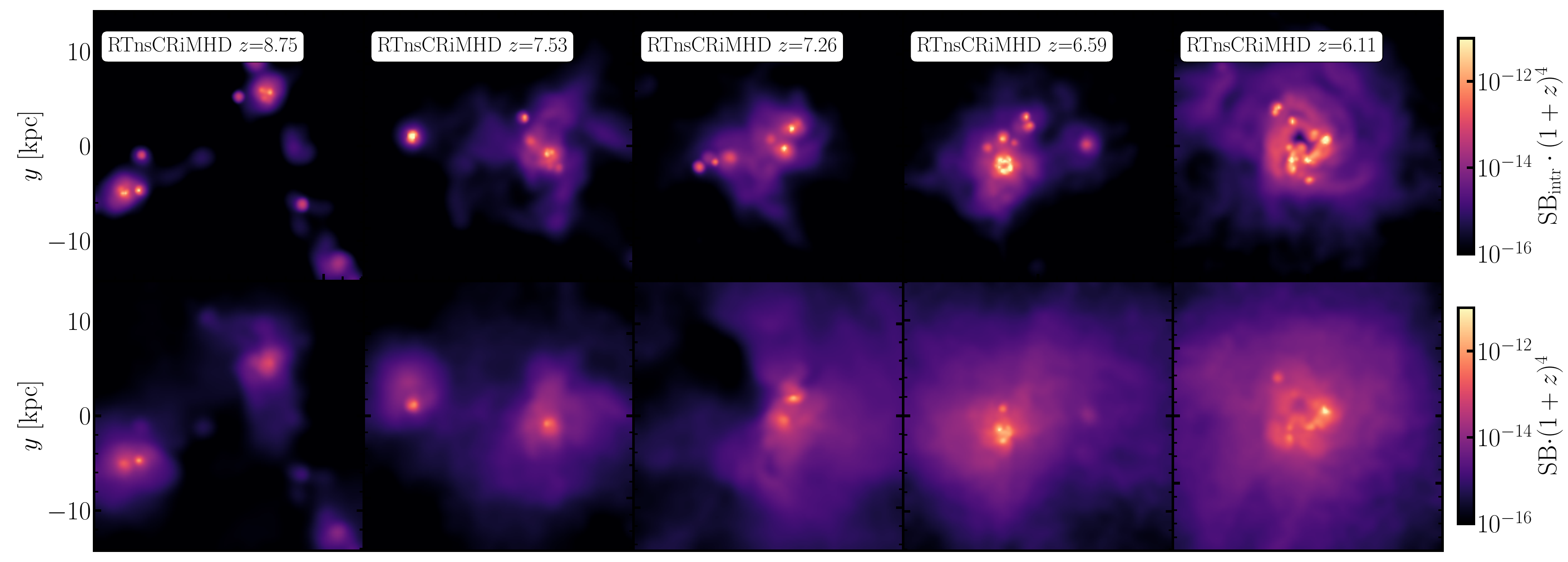}
    \end{subfigure} 
    \begin{subfigure}[b]{.82\textwidth}
     \centering
     \includegraphics[width=\textwidth]{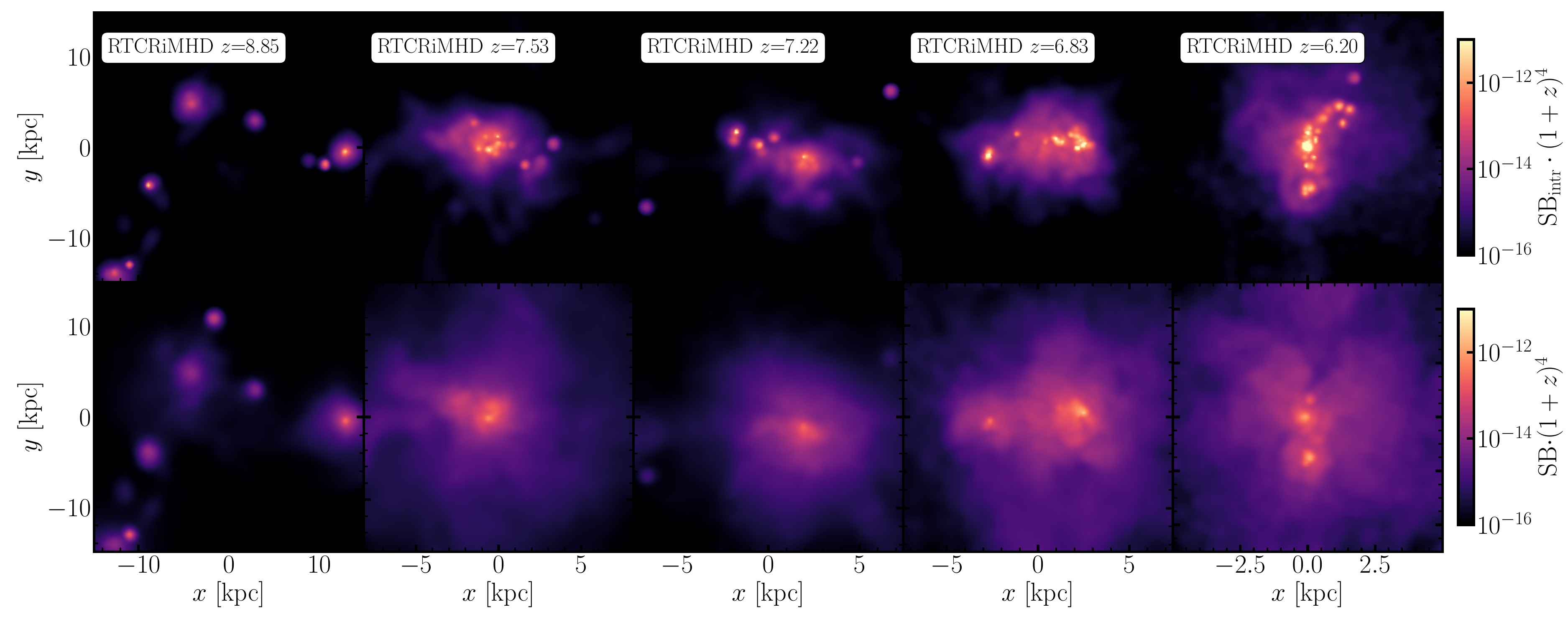}
    \end{subfigure} 
    \caption{Ly$\alpha$ images across the merger sequences. From top to bottom, the four sets (two rows for each set) show the maps for the different simulations. In each set, the first row shows the intrinsic Ly$\alpha$ emission and the second shows the intrinsic scattered Ly$\alpha$ emission. The units of the surface brightness SB are erg s$^{-1}$ cm$^{-2}$ arcsec$^{-2}$. The resonant scattering of the ISM/CGM makes \lya images much more diffuse and the merger signatures much less prominent.
    }
    \label{fig:lya_img_diff_phys}
\end{figure*}

We perform the spectral fitting and analysis for Ly$\alpha$ spectra with the pipeline described in section~2.2.3 of \citetalias{Yuan_Yuxuan24}. We first fit single, double, and triple peak asymmetric Gaussian profiles \citep{Shibuya14_lae_kin_dist} to the \lya spectra. We next decide how many peaks give the best fit based on the Akaike information criteria (AIC; \citealt{Sharma17}). We then extract the following diagnostic quantities of the Ly$\alpha$ spectra according to the fits,
\begin{itemize}
    \item $\vred$, the velocity offset where the maximum of the red peak is reached,
    \item $\sigred$, the velocity width of the red peak,
    \item $A_f$, the asymmetry parameter of the red peak\footnote{This quantity is defined as $A_f = \int_{\vred}^\infty f_v\,dv/\int_{v_{\rm valley}}^{\vred} f_v\,dv$, where $f_v$ is the simulated \lya spectrum and $v_{\rm valley}$ is the velocity of the central valley (see Equation~(33) of \citealt{Kakiichi&Gronke21}).},
    \item and EW, the rest-frame equivalent width.
\end{itemize}

We calculate the intrinsic scattered (by the ISM/CGM) \lya emission\footnote{We note that the term 'intrinsic \lya spectra' in several papers (like \citealt{Mason18_baye_igm, Witstok24_env_eor_lae}) has the same meaning as the term 'intrinsic scattered \lya spectra' used in this paper.} as well as the observed \lya emission (including  IGM attenuation for three transmission levels and line smoothing).

We also calculate \lya quantities assuming Gaussian red peaks. For this we first fit a Gaussian profile to the red peak of our intrinsic scattered spectra. We then convolve the Gaussian fit with the IGM damping curve and account for the line spread function (LSF) in the same way as described in \autoref{ssec:igm_attenuation} and \autoref{asec:model_dw_inst}. We then again calculate the above \lya quantities. Any quantity calculated in this way will be denoted with a subscript $_{\rm Gauss}$\footnote{Note that we investigate the results of several alternative definitions of quantities for which a Gaussian red peak has been assumed in \autoref{ssec:res_lae_xhi}.}.

\section{Results}
\label{sec:res}

In this section, we systematically investigate the effect of the different physical processes. A prominent galaxy merger occurs at $z\approx 7.3$ in all the simulation runs and the runs feature similar evolution sequences, albeit with some differences, as discussed in \autoref{ssec:diff_phys_evol}. We then move on to explore the properties of the Ly$\alpha$ spectra in \autoref{ssec:lya_spec}. Next, we compare our results with recent JWST observations in \autoref{ssec:comp_obs}. Finally, in \autoref{ssec:res_lae_xhi} we point out the implications of our results for two LAE-based inference models of the neutral fraction $\xhi$.

\subsection{The galaxy merger sequence}
\label{ssec:diff_phys_evol}

In this section, we investigate the evolution of galactic properties and Ly$\alpha$/LyC observables from the galaxy merger phase to the quasi-steady disc phase in the redshift range $z \sim 9 \to 6$\footnote{Interested readers may also refer to \aref{asec:add_anal_merger} for additional figures and analysis.}. 

We show the time evolution of the intrinsic and intrinsic scattered Ly$\alpha$ images for four simulations in \autoref{fig:lya_img_diff_phys}. The maps of the intrinsic Ly$\alpha$ emission closely mirror the gas distribution. The maps of the Ly$\alpha$ emission scattered by neutral hydrogen in the ISM and the CGM on the other hand are much more diffuse. This reduces the number of discernible components and makes the merger signatures much less prominent. For example, in the RTnsCRiMHD simulation at $z=7.26$, the number of identifiable components reduces from 6 to 2 between the maps of intrinsic and scattered \lya emission. In the RT and RTiMHD simulations the merger signatures are nearly lost as a result. Accounting for magnetic fields leads to only minor differences. It may increase the clumpiness of the intrinsic \lya images at $z < 7$, although this is clearly washed out in the scattered emission. The inclusion of cosmic rays has a more prominent effect, observed in the two simulations including cosmic rays, for which the Ly$\alpha$ maps of the intrinsic emission are more concentric and the signatures of the ongoing merger are reduced but remain visible, although in most cases the number of identifiable objects also reduces as discussed in \citetalias{Witten24}.

\begin{figure*}
    \centering
    \includegraphics[width=1.\linewidth]{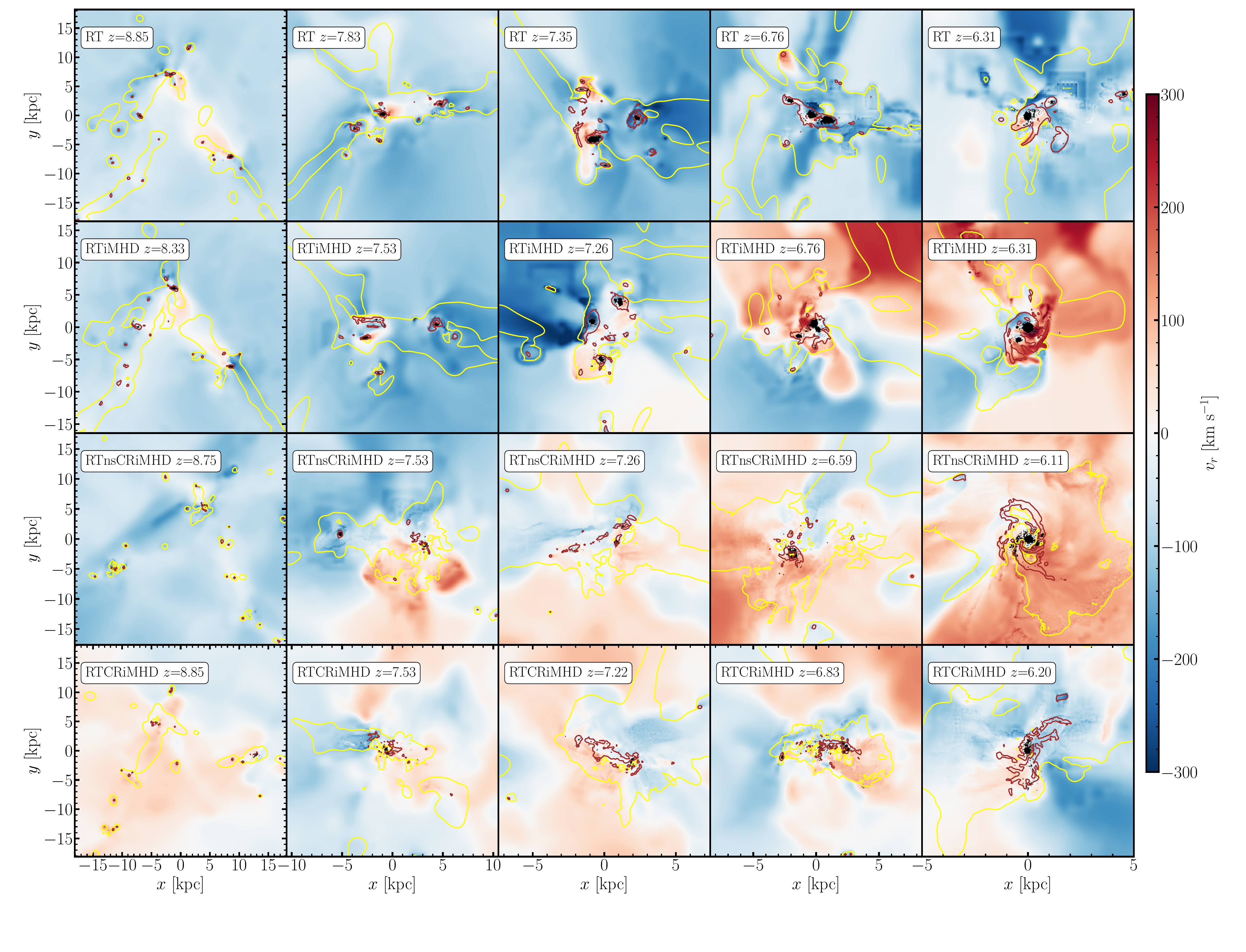}
    \caption{Projected maps of the H~\textsc{i}-mass weighted radial velocity $v_r$ across the merger sequence. Different rows show the results of different simulation runs. The yellow (red) contours represent column densities of $N_{\rm H~\textsc{i}} = 3\times 10^{20}\,\mathrm{cm}^{-2}$ ($10^{22}\,\mathrm{cm}^{-2}$). The black dots represent young star particles with age $t_*<30$~Myr. Most but not all young stars are shielded by high  H~\textsc{i} column densities.}
    \label{fig:img_diff_phys_vr}
\end{figure*}

In \autoref{fig:img_diff_phys_vr}, we show maps of the projected H~\textsc{i} mass-weighted radial velocity $v_r$ along the merger sequence\footnote{The velocity has been subtracted by the mass-weighted mean velocity of the merger system.}. Outflows powered by merger-driven star formation show up in these maps as red regions. 
As investigated in \citet{Rodriguez-Montero24} and in \citet{Martin-Alvarez25} with a similar physical configuration and solver to ours, simulations including cosmic rays are much more efficient in the regulation of star formation rates and in driving efficient outflows capable of escaping from the galaxy. This naturally leads to higher mass loading factors, and to outflows more multi-phase in nature that entrain a higher proportion of high temperature gas. As we will discuss later, this has a major effect on the \lya emission due to the clearing of escape channels for Lyman continuum and \lya radiation \citepalias{Yuan_Yuxuan24}. The more efficient galactic winds also lead to a very different dust distribution and much less absorption of the \lya emission by dust. Most but not all recent star formation is shielded by high columns of neutral hydrogen. Note also that during the pre-merger phase, the surrounding CGM densities of the triple galaxy system in RTnsCRiMHD and RTCRiMHD are much lower than those of RT and RTiMHD, possibly due to stronger outflows pushing the gas out. During the merger phase ($z \sim 7.2$), we see much stronger and more continuous outflows in many directions out to much larger radii in the two simulations with cosmic rays, compared to the more gentle galactic winds in the RT and RTiMHD simulations. The stronger feedback also leads to a more spread-out spatial distribution of young stars and creates lower-density channels. Furthermore, the metals (and thus dust) are carried out to much larger radii by the outflows in the two simulations including cosmic rays, while the metals are significantly more spatially concentrated in the RT and RTiMHD simulations (see also \autoref{fig:img_diff_phys_Z} in the \aref{asec:add_anal_merger}). Beyond the more dominant effects resulting from the inclusion of cosmic rays, some additional effects are also apparent. Accounting for magnetic fields drives higher and somewhat faster outflows, as these increase star formation clustering in our prescription through a higher non-thermal support against small-scale collapse  \citep{Martin-Alvarez20}. Including cosmic ray streaming and the resulting heating seemingly reduces outflow velocities, as these smooth CR gradients and drive denser outflows \citep[e.g.,][]{Wiener17}.

\begin{figure*}
    \centering
    \includegraphics[width=0.9\linewidth]{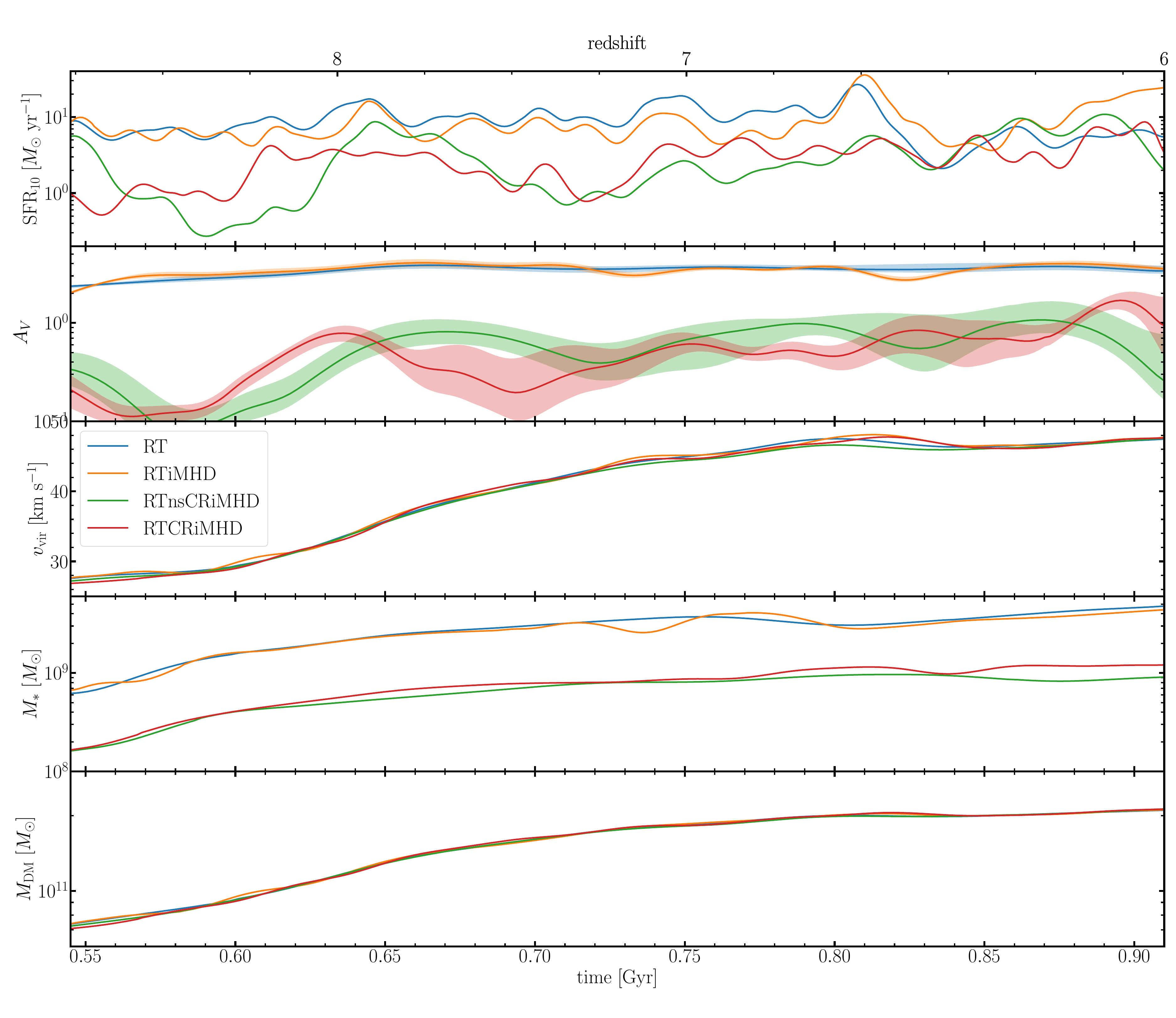}
    \caption{Time evolution of the star formation rate averaged over $10$~Myr (SFR$_{10}$), $A_V$, $v_{\rm vir}$, $M_*$, $M_{\rm DM}$. Different colours represent different simulation runs. In the $A_V$ panel, the shaded region represents the 1$\sigma$ scatter when evaluated along different LOS.}
    \label{fig:gas_time_evo_diff_phys}
\end{figure*}

\begin{figure*}
    \centering
    \includegraphics[width=0.9\linewidth]{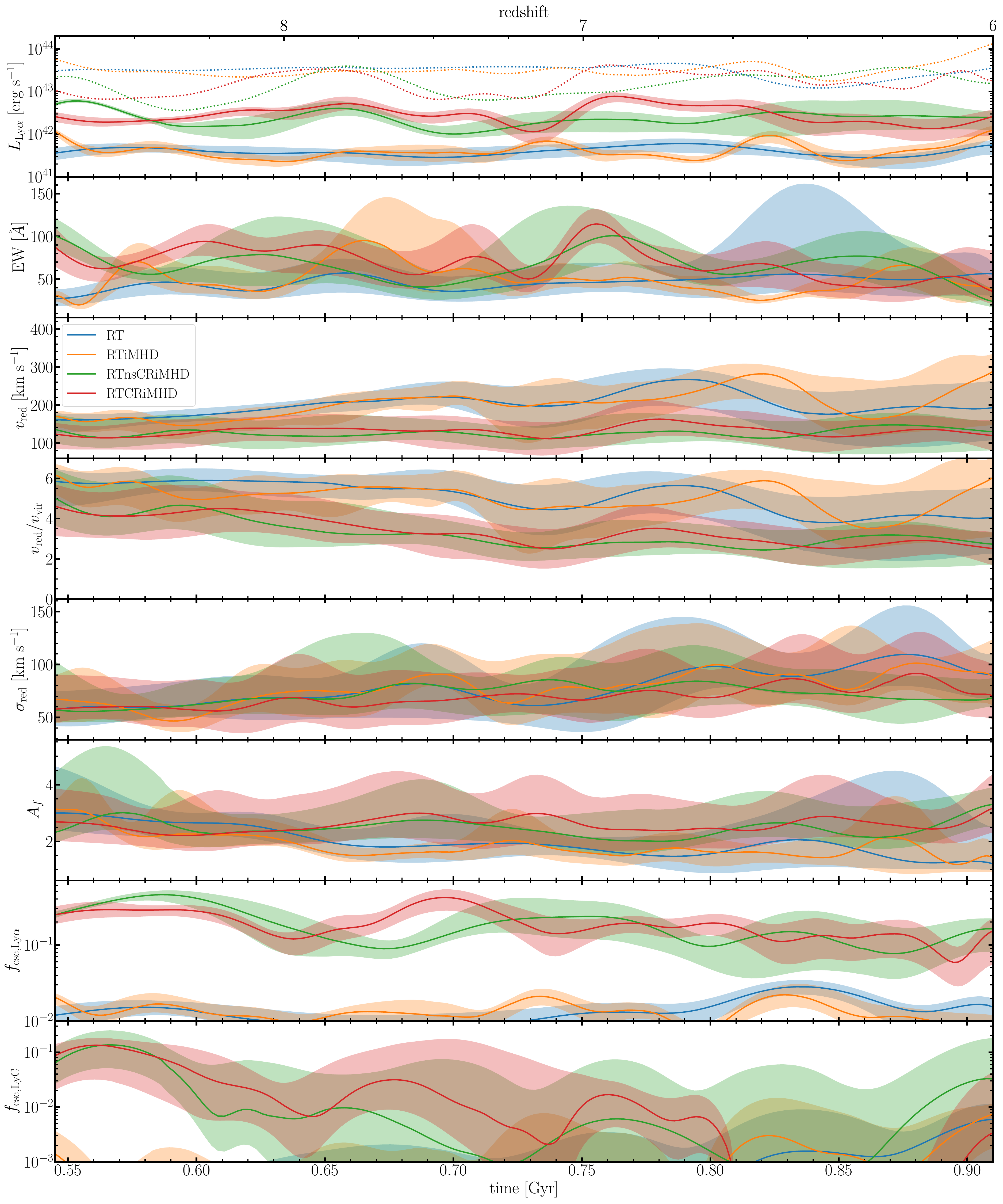}
    \caption{Time evolution of $L_{\rm Ly\alpha}$, EW, $\vred$, $\vred/v_{\rm vir}$, $\sigma_{\rm red}$, $A_f$, $\fesclya$, and $\fesclyc$. Different colours represent different simulation runs. In the $L_{\rm Ly\alpha}$ panel, the dotted curves show the value of intrinsic $L_{\rm Ly\alpha}$ (without ISM/CGM scattering). For $L_{\rm Ly\alpha}$, EW, $\vred$, $\vred/v_{\rm vir}$, $\sigma_{\rm red}$, $A_f$, $\fesclya$, and $\fesclyc$, the solid curves with the shaded region show the mean intrinsic scattered values with their $1\sigma$ scatter when evaluated across different LOS. 
    }
    \label{fig:lya_time_evo_diff_phys}
\end{figure*}

In \autoref{fig:gas_time_evo_diff_phys} we show the time evolution of several physical properties of the triple galaxy system for all simulation runs\footnote{In the merger phase, the calculations of $\vvir$, $M_*$, and $M_{\rm DM}$ are for the main halo hosting all three galaxies.}. In both the RT and RTiMHD simulations the star formation history (SFH) is relatively smooth at a level of $\sim 10\,\Msun\,\yr^{-1}$, whereas the RTnsCRiMHD and RTCRiMHD simulations have more bursty SFHs before $z \sim 7$ and star formation becomes more steady after the last merger event at $z \sim 7$ \citep{Dome24}. The average star formation rates (SFRs) in the simulations including cosmic rays are significantly lower, reflecting the more effective feedback due to stronger galactic winds. Both the RT and RTiMHD simulations have a smooth angular distribution of $A_V$ with rather large values of $A_V \sim 3$, reflecting the larger metallicity produced by the much higher star formation rates and the rather compact distribution of the dust due to the rather inefficient galactic winds. In the simulations including cosmic rays the angular distribution of $A_V$ is highly anisotropic and varies rapidly with time along the merger sequence with low values varying from 0.1 to 1, comparable to those of observed LAEs \citep{Simmonds23, Tang_Mengtao23, Witstok25_pri_lae}. The virial velocity $\vvir$ of the halo hosting the galaxy mergers increases monotonically in all the simulations. The stellar mass steadily grows in all simulation runs and increases from $10^9 \Msun$ to $5\times 10^9 \Msun$ ($2\times 10^8 \Msun$ to $2\times 10^9 \Msun$) in the RT/RTiMHD (RTnsCRiMHD/RTCRiMHD) simulation. The inclusion of magnetic fields appears to have some minor effect on the SFR and stellar mass, although this remains secondary and predominantly affects other galactic properties \citep[e.g.,][]{Martin-Alvarez20, Rodriguez-Montero24, Robinson24}. The dark matter (DM) mass of the halo hosting the merger is pretty much constant with a modest steady increase with time, and the host halo of the final post-merger disc galaxy has a mass around $2\times 10^{11} \Msun$ in all simulations. 

We show the time evolution of the properties of the Ly$\alpha$ and LyC emission in \autoref{fig:lya_time_evo_diff_phys}. The Ly$\alpha$ and LyC observables show large LOS variations \citep{Barnes11}, so we show the median and the range for all LOSs. The time evolution of the intrinsic  Ly$\alpha$ luminosity, $L_{\rm Ly\alpha}$, closely follows that of the SFR. The evolution of the luminosity of the intrinsic scattered \lya emission is similar to that of the intrinsic \lya emission because of the less episodic evolution of $\fesclya$. However, we see that the RT and RTiMHD simulations have significantly lower intrinsic scattered $L_{\rm Ly\alpha}$  due to their much lower $\fesclya$ because of the high $A_V$ in these two simulations. The evolution of EW and width of the red peak $\sigma_{\rm red}$ are episodic for all simulation runs, with EW ($\sigma_{\rm red}$) varying from $30-150$~\AA\, ($50-150$~km~s$^{-1}$). The $\vred$ evolution of the RT and RTiMHD simulations is generally more episodic and reaches high values of $\sim 250$~km~s$^{-1}$, while those of the simulations including cosmic rays are relatively constant over the merger sequence and have lower values of $\sim 150$~km~s$^{-1}$. We further show the ratio of $\vred$ to the virial velocity $v_{\rm vir}$ and see that its evolution has a similar trend as that of  $\vred$, with typical values of $\sim 4$ for the RT and RTiMHD simulations and $\sim 2$ for the two simulations including cosmic rays. The \lya spectra in the  RT/RTiMHD (RTnsCRiMHD/RTCRiMHD) simulations have an asymmetry factor of $A_f \sim 2 (3)$. The asymmetry is due to Ly$\alpha$ photons escaping both by resonant scattering to the line wings and through the low-density channels \citep{Kakiichi&Gronke21, Kimm22}.  The RT/RTiMHD simulations show a higher level of symmetry because there are fewer low-density channels and \lya photons mainly escape by scattering to the line wings. The RT/RTiMHD simulations have very low $\fesclya$ and $\fesclyc$ because of their high dust contents (described below) while the simulation runs including cosmic rays have $\fesclya$ and $\fesclyc$ varying around $\sim 0.2$ and $\sim 0.1$. For the two CR runs, $\fesclya$ is well synchronized with the EW. 

In summary, the results discussed in this section illustrate the important effect that magnetic-field-mediated cosmic ray diffusion may have in modulating the reionization process (compare RTiMHD to RTnsCRiMHD) and we see that CR streaming generally plays a much smaller role than CR diffusion (compare RTnsCRiMHD to RTCRiMHD)\footnote{This phenomenon might happen because the adopted constant CR diffusion coefficient is relatively high \citep{Quataert22_cr_streaming}.}. Our results also confirm that the role of pure magnetic fields (without CR) remains secondary for the escape of ionizing radiation (compare RT to RTiMHD), except when extreme magnetisations are considered \citep{Katz19b, Katz21}.

\subsection{Unattenuated intrinsic scattered Ly\texorpdfstring{$\alpha$}{alpha} spectra}
\label{ssec:lya_spec}

\begin{figure*}
    \centering
    \includegraphics[width=0.9\linewidth]{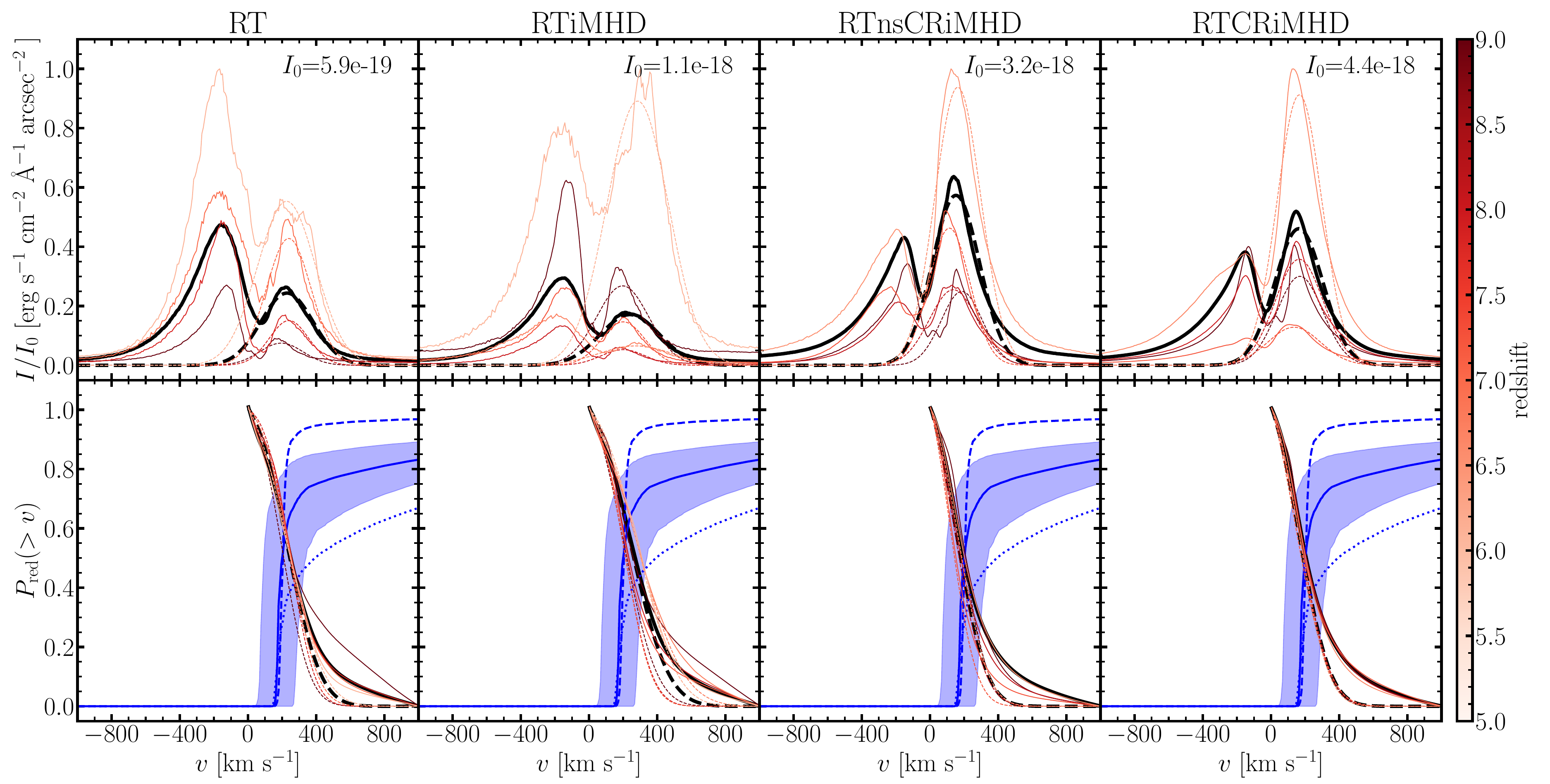}
    \caption{Top row: normalized simulated intrinsic scattered Ly$\alpha$ spectra, where $I_0$ is the normalization constant. Bottom row: cumulative distribution function (CDF) of the flux in the red peak of the Ly$\alpha$ spectra shown above. From left to right we show the results of different simulation runs. The solid coloured curves show the angle-averaged spectra at different redshifts while the solid black lines show both angle- and time-averaged spectra. Redder colour represents higher redshifts (as shown by the colourbar). The dashed curves show single Gaussian fits to the red peaks. In the bottom row, we also overplot the median damping wing curve at $z=6,7,8$ as dashed, solid and dotted blue lines, respectively. The shaded regions show the 1$\sigma$ uncertainties associated with the $z=7$ IGM damping wings. In all simulation runs, the tails of the red peaks extend to higher velocities than predicted by the Gaussian fits.
    }
    \label{fig:spec_sum}
\end{figure*}

We plot the unattenuated intrinsic scattered \lya spectra in the top row of \autoref{fig:spec_sum}. Our simulations reproduce a diversity of \lya line shapes over the merger sequence for the four simulated physics implementations, qualitatively consistent with the results of \citet{Blaizot23}\footnote{On a first glimpse, it looks as if our results show a larger fraction of emission in the blue peak (larger blue-to-total ratio B/T) than low-$z$ observations \citep{Steidel10, Kulas12, Erb14, Rivera-Thorsen15, Gronke16, Verhamme17, Naidu22a}. There are three possible reasons for this: (1) \citet{Blaizot23} found that the B/T ratio anti-correlates with \lya luminosity and EW. There is thus likely an observational bias towards spectra with low B/T (higher luminosity); (2) High-$z$ \lya spectra may have larger values of B/T than those at low-$z$, possibly due to stronger inflow dynamics \citep{Vitte25} . However, several works also report non-evolution of \lya spectra between $z\sim 2-6$ \citep{Santos20, Hayes21, Hayes23c}; (3) The lack of hydro-resolution for resolving low-density channels may predict stronger blue wing emission, as discussed in \autoref{ssec:caveats}.}. To investigate the symmetry of the red peak, we fit single Gaussians to the red peaks of the spectra and show the corresponding results as dashed curves. We see that in all simulation runs the red peaks are asymmetric and deviate from a Gaussian in both centre and line wing and the tails of the red peaks extend to high velocities. This effect is more evident in the simulations including cosmic rays, consistent with their higher values of the asymmetry factor $A_f$ (see \autoref{fig:lya_time_evo_diff_phys}). In order to see the contribution from the non-Gaussian tail more clearly, we compare the cumulative distribution function (CDF) of the red peak to the damping wing curves at $z=6,7,8$ in the bottom row of \autoref{fig:spec_sum}. We see that the neutral hydrogen in the IGM allows a high transmission fraction beyond $\sim 400$~km~s$^{-1}$. In our fiducial RTCRiMHD simulation, the tails beyond $400$~km~s$^{-1}$ contribute $\sim 20\%$ of the emission in the red peak, while if we use Gaussian fits they only contribute $\sim 5\%$. Hence we conclude that modelling the red peak of the intrinsic scattered spectrum as a Gaussian will significantly underestimate the transmission of \lya at high redshift when the IGM is still significantly neutral\footnote{Although not investigated in this paper, we should note that the assumption of Gaussian profiles also has a strong impact on the \lya luminosity function (LF) of the high-$z$ LAE population \citep{Garel21}.}. For the RT~simulation this effect is particularly pronounced at higher redshift, but the inclusion of magnetic fields preserves this to lower redshifts. The inclusion of the additional non-thermal physical processes associated with cosmic rays leads to a significantly more pronounced effect with time, which is maximised in the model accounting for CR streaming. As we will discuss below, this becomes even more important if the rather uncertain amount of dust in the simulation is reduced \citep{Xu_C23}. 

\begin{figure*}
    \centering
    \includegraphics[width=.7\linewidth]{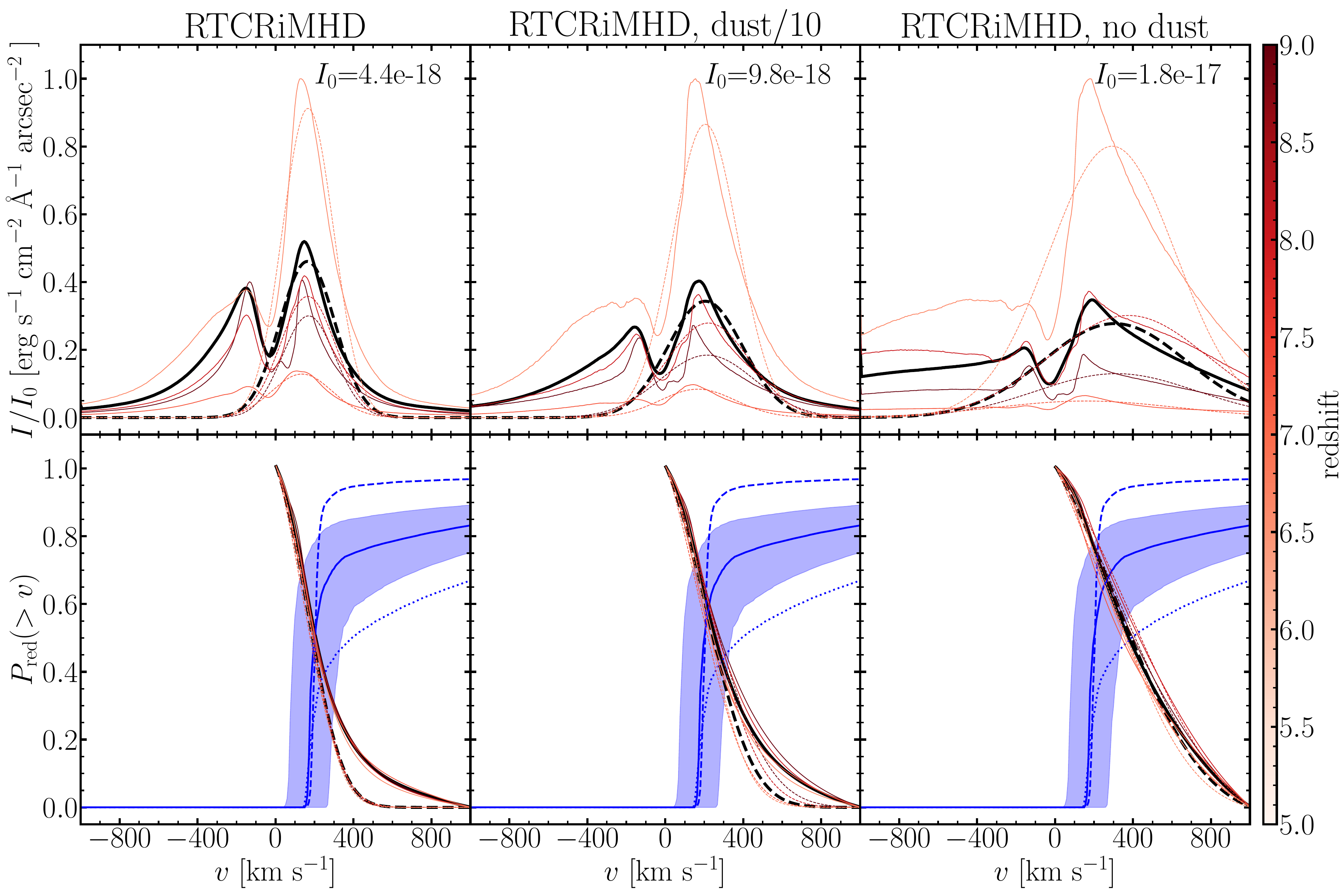}
    \caption{Same as \autoref{fig:spec_sum}, but from left to right we show the results for the RTCRiMHD simulation with three levels of dust contents: fiducial, dust content reduced by a factor of 10 (dust/10), and removing all dust (no dust). With decreasing dust content the profiles extend significantly further to the red.}
    \label{fig:spec_sum_diff_dust}
\end{figure*}

We next investigate the impact of dust on the shape of the \lya spectra in our fiducial RTCRiMHD simulation and show the results in \autoref{fig:spec_sum_diff_dust}. From the intrinsic scattered spectra shown in the top row, we see that as the dust content decreases, the \lya emission becomes more luminous ($I_0$ increases). Note that due to the resonant scattering of \lya this effect is highly non-linear. In addition, the spectra also become much more asymmetric with much more extended red wings\footnote{We note here that we cannot easily quantify the increased levels of asymmetries by comparing the CDF of simulated \lya red peaks and single Gaussian fits. This is because for the cases of ``dust/10'' and ``no dust'', the \lya profiles are highly asymmetric and single Gaussian profiles generally give a bad fit.}. 
The \lya spectra extend to much larger velocities, as can be seen more clearly from the CDF shown in the bottom row, where we find that the tails beyond $400$~km~s$^{-1}$ contribute $\sim 20\%$, 30\%, 50\% for the ``fiducial'', ``dust/10'', and ``no dust'' cases, respectively. The physical effects behind this are two-fold. The first effect is that as the dust content decreases, \lya photons are less likely to be absorbed, undergo many more resonant scattering events, and diffuse in velocity space. The second effect is that they reach larger radii where strong inflows/outflows co-exist, and get scattered by this high-velocity gas. 

\begin{figure*}
    \centering
    \includegraphics[width=.9\linewidth]{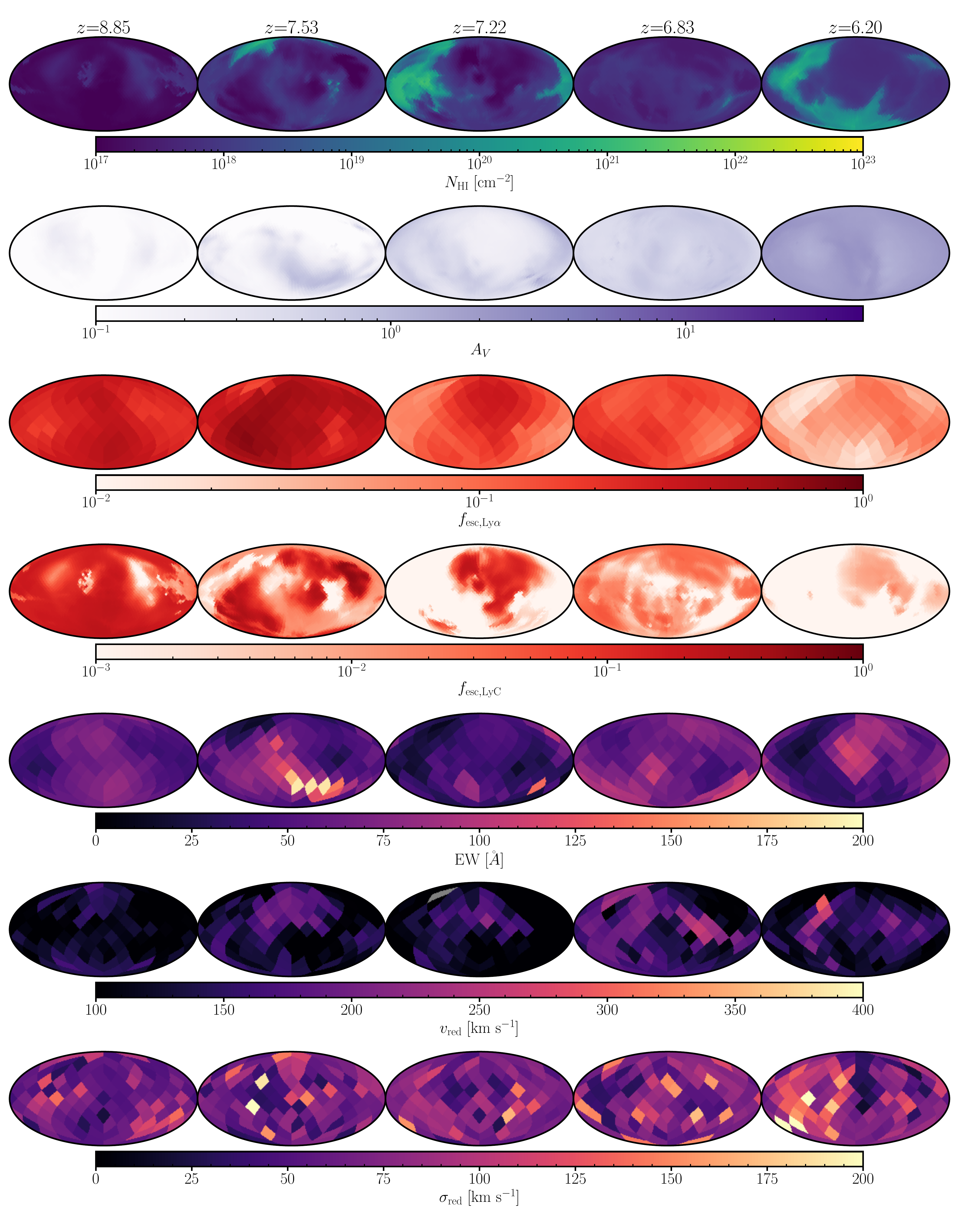}
    \caption{Angular distribution of $\NHI$ and $A_V$, $\fesclya$, $\fesclyc$, EW, $\vred$, and $\sigma_{\rm red}$ across the merger sequence of the RTCRiMHD simulation. $\fesclya$, $\fesclyc$, EW, $\vred$, and $\sigma_{\rm red}$ are for emission that has emerged from the galaxies.}
    \label{fig:ang_dist_diff_time}
\end{figure*}

We next investigate the variations due to different LOSs. For this, we show the angular distribution of various gas and \lya parameters across the merger sequence of the RTCRiMHD simulation that we consider the most realistic in \autoref{fig:ang_dist_diff_time}. Both \lya and LyC photons escape through low-density channels visible in the $\NHI$ panels. $A_V$ shows a smoother angular distribution compared to $\NHI$, indicating that metals are spread out efficiently by outflows. In the disc phase at $z\sim 6.2$,  $\fesclya$ and $\fesclyc$ show a bipolar angular distribution in the simulations including cosmic ray-driven outflows, suggesting both Ly$\alpha$ and LyC preferentially escape in a direction perpendicular to the disc galaxy and along the rotation axis, consistent with the findings of \citet{Verhamme12, Behrens&Braun14, Costa22}. $\fesclya$ thereby has a positive correlation with EW and a negative correlation with $\vred$. $\sigred$ does not show any clear correlation with other quantities.

\begin{figure*}
    \centering
    \includegraphics[width=0.8\linewidth]{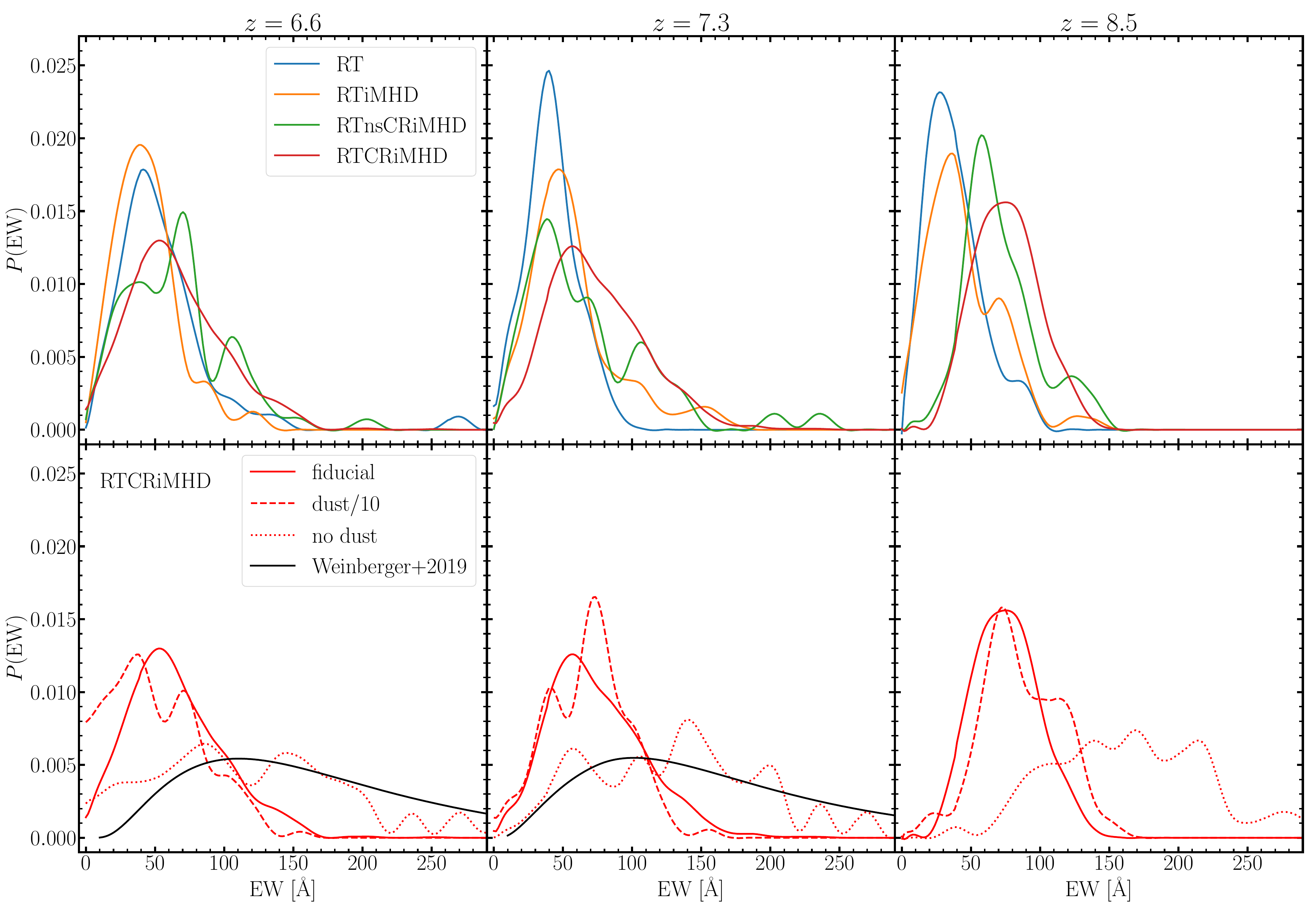}
    \caption{EW distribution of the intrinsic scattered \lya emission at $z=6.6, 7.3, 8.5$ (from left to right). The top row shows the effect of different simulated physics represented by lines of different colours. The bottom row shows the effect of different dust contents for the RTCRiMHD simulation. We also overplot the intrinsic scattered EW distribution from \citet{Weinberger_L19}, based on the EW distribution originally proposed by \citet{Dijkstra&Wyithe12}. 
    }
    \label{fig:EW_dist}
\end{figure*}

We show the EW distribution of the intrinsic scattered \lya emission for the simulations with different physics implementations at $z= (6.6, 7.3, 8.5) $ in the top row of \autoref{fig:EW_dist}. We see that the inclusion of magnetic fields leads to only minor differences, and most apparent at $z \gtrsim 7$, broadening the EW distributions to higher values. We find cosmic rays to have the most important impact, extending the distribution to larger EWs, and displacing its peak to $75{-}100$~\AA, especially at the highest redshift we consider, $z\sim 8.5$. As before, accounting for CR streaming appears to enhance this effect in a secondary, yet non-negligible manner. The EWs appear somewhat smaller and the distribution appears to be narrower than the distribution proposed by \citet{Dijkstra&Wyithe12} and \citet{Weinberger_L19} for the intrinsic scattered emission.  We investigate the impact of dust for our fiducial 
RTCRiMHD in the bottom row of \autoref{fig:EW_dist}. Removing all the dust content makes the distributions broader and extend to larger EWs. Dust thus plays, as expected, an important role in regulating the EW also at those redshifts. Interestingly, there is good agreement with the distribution of  \citet{Weinberger_L19} when we assume that there is no dust. We lastly caution that a more top-heavy IMF and burstier SF would shift the intrinsic scattered EWs to larger values \citep{Charlot&Fall93, Garel15}.

\subsection{Comparing to observations of high-\texorpdfstring{$z$}{z} LAEs}
\label{ssec:comp_obs}

In this section, we compare the properties of our simulated LAE to the observations of high-$z$ LAE populations. 

\begin{figure*}
    \centering
    \includegraphics[width=0.9\linewidth]{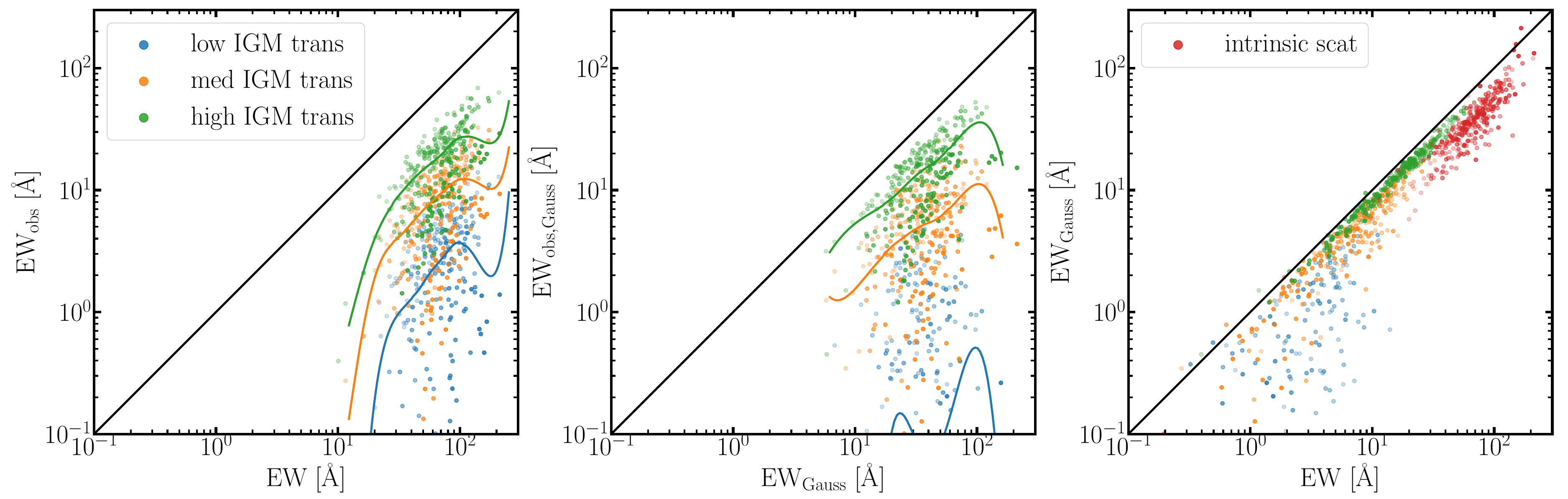}
    \caption{Left panel: observed IGM-attenuated EW versus intrinsic scattered EW. Middle panel: observed IGM-attenuated EW versus intrinsic scattered EW, but both quantities are calculated with the assumption of a Gaussian profile. Right panel: EW$_{\rm Gauss}$ versus EW, for both observed IGM-attenuated and intrinsic scattered values (shown by red dots) for the RTCRiMHD simulation. Different colours represent different IGM transmission levels. Higher opacities of markers represent higher redshift. In the left two panels, we show the median relations for the data points of different IGM transmission levels. Assuming Gaussian profiles increasingly underestimates  EWs with decreasing transmission levels.}
    \label{fig:ew}
\end{figure*}

We compare EW$_{\rm obs}$ of the observed \lya emission with the EW of the intrinsic scattered \lya emission in the left panel of \autoref{fig:ew}. As expected,  EW$_{\rm obs}$ is systematically lower than EW due to the IGM attenuation. There is a larger spread of EW$_{\rm obs}$ at low IGM transmission, as only the red tails of the \lya spectra are transmitted. We compare EW$_{\rm obs}$ versus EW, but both are now calculated with the Gaussian fits, in the middle panel. We see similar trends as before, but the reduction of the EW is more severe at low IGM transmission. We finally compare EW$_{\rm Gauss}$ versus EW. We see that EW$_{\rm Gauss}$ is always lower than the full EW, as the Gaussian fits do not have the extended line wings. The extent of the underestimation again increases with decreasing IGM transmission levels.

\begin{figure*}
    \centering
    \includegraphics[width=0.9\linewidth]{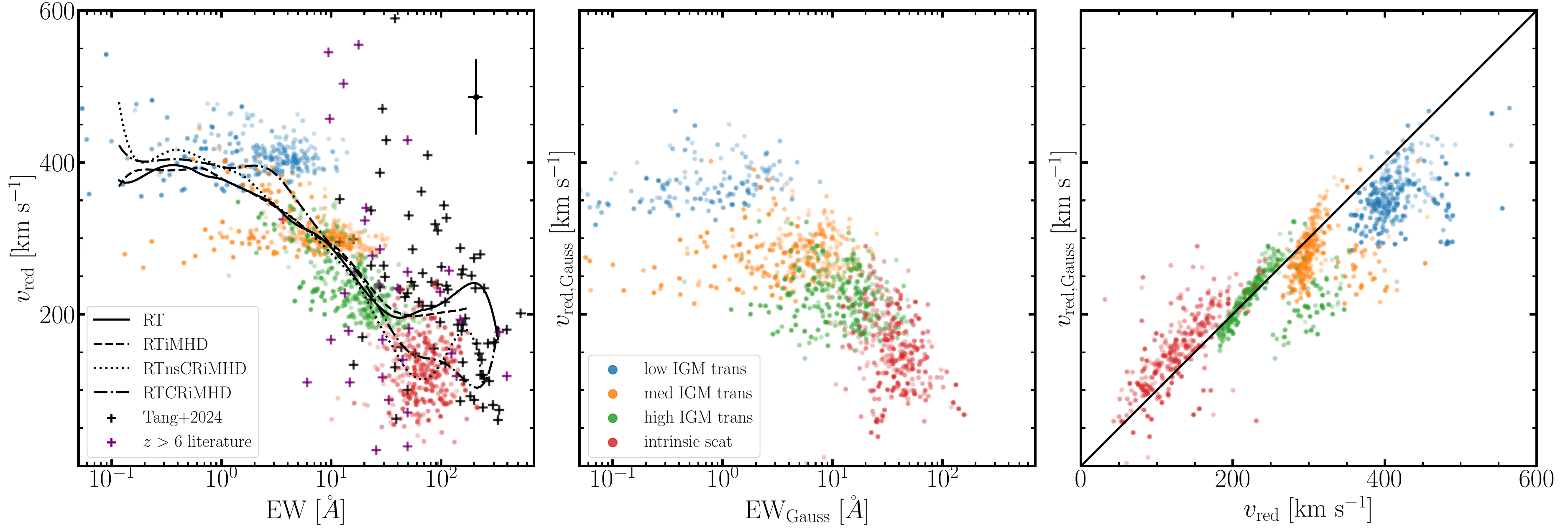}
    \caption{Left panel: $\vred$ versus EW. Middle panel: $\vred$ versus EW, but both quantities are calculated with the assumption of a Gaussian profile. Right panel: $v_{\rm red, Gauss}$ versus $\vred$ for the RTCRiMHD simulation. Different colours represent different IGM transmission levels. Higher opacities of markers represent higher redshift. In the left panel, the black lines of different line styles represent the median relations for different simulation runs. In the left panel we also overplot the literature data of observed $z\sim 5-6$ LAEs as black crosses (\citealt{Tang_Mengtao24_lya_z5_6}; with representative error bar on the top right corner), and $z>6$ galaxies as purple crosses \citep{Maiolino15, Stark15, Stark17, Willott15_lbg, Inoue16, Pentericci16, Bradak17, Laporte17, Mainali17, Hashimoto19, Hutchison19, Endsley22, Bunker23, Saxena24}. The $\vred$ of the simulated spectra show a strong anti-correlation with the EWs due to the effect of IGM attenuation.}
    \label{fig:vred_ew}
\end{figure*}

In the left panel of \autoref{fig:vred_ew} we plot the velocity of the red peak $\vred$ versus EW. The different IGM transmission levels are driving an anticorrelation between these two quantities, as observed in \citet{Saxena23}  and \citet{Tang_Mengtao24_lya_z5_6}. The physical reason is simple: with higher IGM transmission, the observable flux in the IGM-attenuated \lya spectra extends further to the blue. Different lines in this panel display the median relation for each of the studied models. These reflect the same general pattern: inclusion of CRs leads to the most pronounced differences, although the differences between the median distributions of the models are relatively minor. In the middle panel we plot  $\vred$ versus EW  with both quantities calculated from Gaussian fits. The trend of an anti-correlation is preserved, but there is a larger amount of scatter. In the right panel, we compare $\vred$ calculated with Gaussian fits versus the true values. We see that these two quantities are similar for the intrinsic scattered  emission without IGM attenuation and the case of high IGM transmission. However, at low IGM transmission, $v_{\rm red, Gauss}$ is much lower than $\vred$, as the unattenuated tails of the spectra extend to much lower velocities for the Gaussian case. Interestingly, we see that all four simulation runs have similar median relations, showing that this relation is robust against including  different physical processes.

\begin{figure*}
    \centering
    \includegraphics[width=0.85\linewidth]{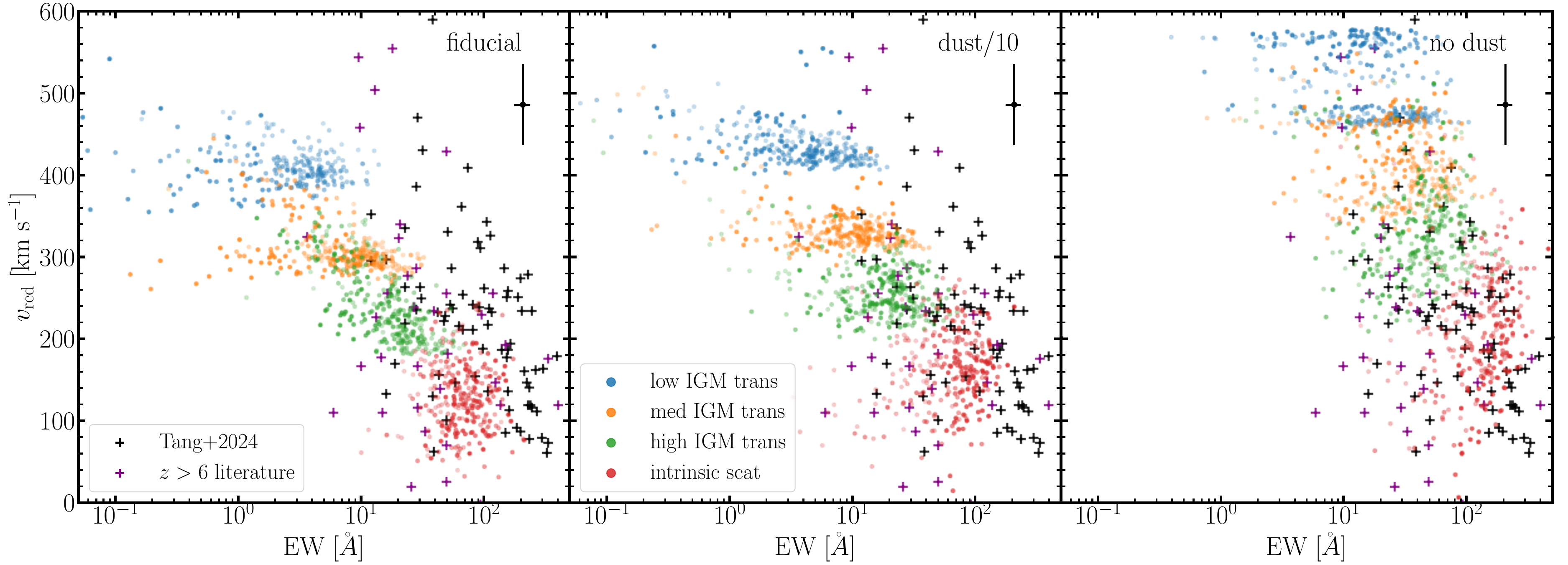}
    \caption{$\vred$ versus EW, but showing different dust contents as indicated by the legend, for the RTCRiMHD simulation. Different colours represent different IGM transmission levels. Higher opacities of markers represent higher redshift. We overplot the same set of observational data \citep{Tang_Mengtao24_lya_z5_6} as in the left panel of \autoref{fig:vred_ew}. Our simulated results assuming no dust agree well with the observations.}
    \label{fig:vred_ew_diff_dust}
\end{figure*}

As we have already discussed in \autoref{ssec:lya_spec}, the amount of dust has a large effect on the \lya spectra. For this reason we investigate the impact of different dust contents on the $\vred$ - EW relation in \autoref{fig:vred_ew_diff_dust}. We see that as the dust content decreases, the simulated data moves in the direction towards the upper right. Interestingly, our simulated data reach good agreement with observations when all the dust is  removed. We should note however again that our simulated results are just from one single galaxy merger, while the observation are from a sample of galaxies. Nevertheless, we can infer  that the impact of dust content might even be larger than the effects of time evolution, LOS variations, and IGM transmission levels. This indicates that constraining dust evolution at early epochs is of crucial importance for interpreting the observations of high-$z$ LAEs. 

\begin{figure*}
    \centering
    \includegraphics[width=0.85\linewidth]{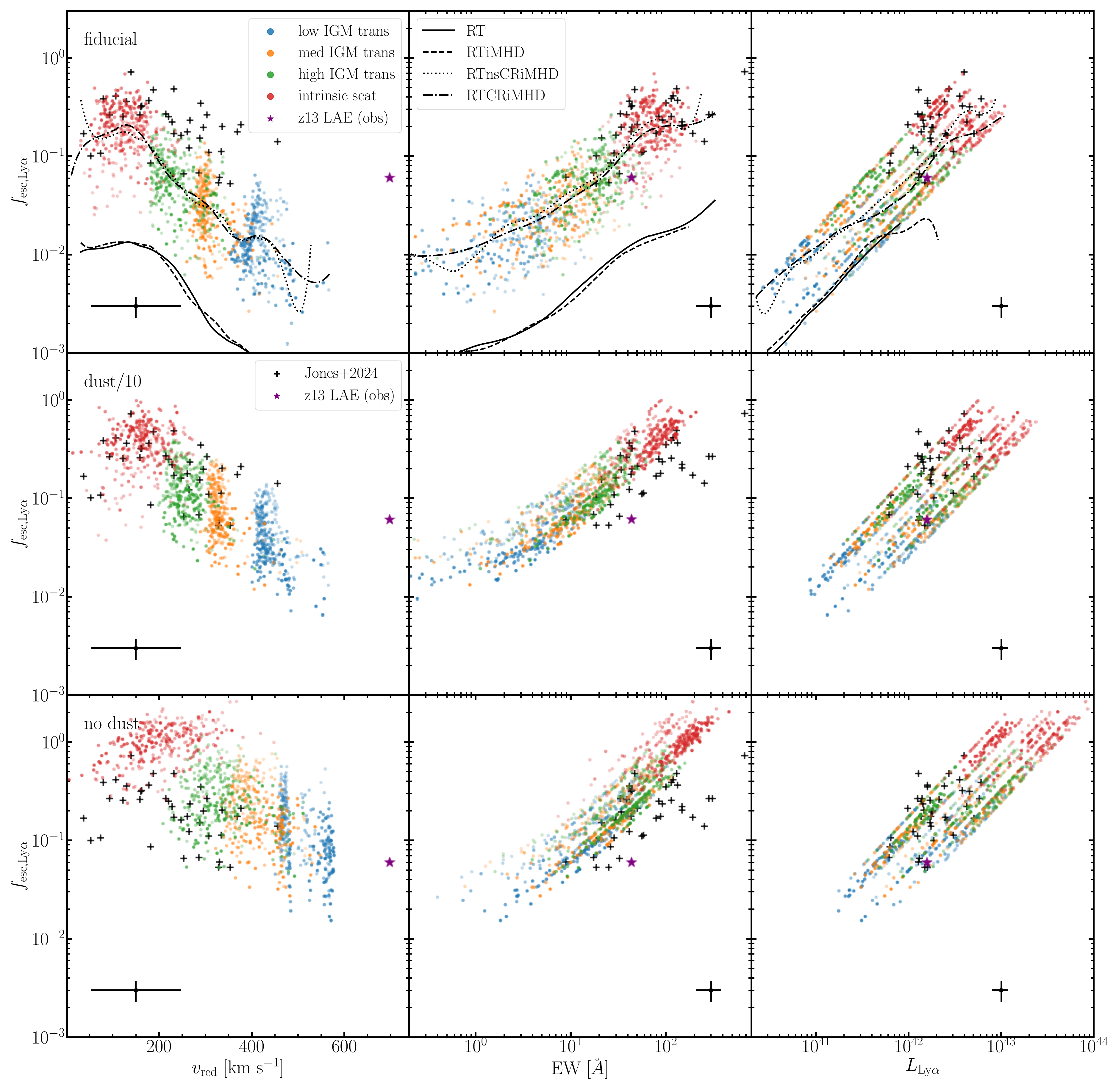}
    \caption{From left to right, we show $\fesclya$ versus $\vred$, EW and $L_{\rm Ly\alpha}$. From top to bottom, we show the results for the fiducial dust model, with dust content decreased by a factor of 10, and with no dust content. The coloured data points are from the RTCRiMHD simulation. Different colours represent different IGM transmission levels. Higher opacities of markers represent higher redshift. In the top rows, the black lines of different line styles represent the median relations for different simulation runs. We also overplot the observational data from \citet{Jones25} as black crosses (in the left column, a representative errorbar is shown on the bottom left, while in the middle and right columns it is shown on the bottom right) as well as the observational data of the recently discovered LAE JADES-GS-z13-1-LA at $z\sim 13$ \citep{Witstok24_lae_z13} as purple star (error bar not shown). Note that the agreement of simulated and observed spectra improves if the assumed dust content is reduced. Note further that the $\vred$ of the $z\sim 13$ LAE has a large uncertainty of $\sim 300$~km~s$^{-1}$, thus our simulated data falls near the lower edge of the error bar of this object.
    }
    \label{fig:deci_fesc_lya}
\end{figure*}

In \autoref{fig:deci_fesc_lya}, we plot the \lya escape fraction $\fesclya$ versus $\vred$, EW and $L_{\rm Ly\alpha}$. We also plot the observational data from \citet{Jones25} as black crosses. The simulations reproduce the negative correlation between $\vred$ and $\fesclya$ \citep{Saxena24}, as well as the positive EW - $\fesclya$ correlation \citep{Saxena24, Tang_Mengtao24_lya_z5_6}. Consistent with our findings earlier, the IGM transmission levels from patchy reionization are the main driver of these correlations in the simulated spectra. We also find a positive correlation between $\fesclya$ and $\LLya$. Comparing the median relations for different simulated physical processes, we find that the trend is maintained for all simulation runs across all three relations. However, the RT/RTiMHD simulations have lower $\fesclya$ by $\sim 1$~dex compared to the two simulations including cosmic rays. The lower dust content results in higher $\vred$, EW, $L_{\rm Ly\alpha}$, and $\fesclya$, hence shifting all three of the $\fesclya-\vred$, $\fesclya$-EW and $\fesclya-\LLya$ relations towards the upper right direction. We see a better agreement with the observational data when the dust content is reduced by a factor of 10, which increases $\fesclya$ by a factor of 3\footnote{Note that due to the resonant nature of \lya scattering we get this increase of the escape fraction despite the already rather low values of $A_V$ of 0.1-0.5 with our fiducial dust content in the RTCRiMHD simulation.}. If we remove the dust altogether, the escape fractions significantly increase further and overshoot the observational data\footnote{For the no dust case of the RTCRiMHD simulation, we had seen that the distribution of intrinsic scattered EW agrees with that of \citet{Dijkstra&Wyithe12} and \citet{Weinberger_L19}. The unattenuated EW distribution observed at low redshift appears thus to predict systematically lower EW values than observed by \citet{Jones25} at high redshift suggesting that the unattenuated EW distribution of the \citet{Jones25} LAE sample is quite different from  the unattenuated EW distribution proposed by \citet{Dijkstra&Wyithe12} due to different selection effects. }. Despite this, the inclusion of cosmic rays leads to variations comparable or larger than removing all dust, except perhaps for the $\fesclya-\LLya$ relation. We again caution that the predicted EWs also depend sensitively on the assumptions of the IMF and other details of the stellar population synthesis modelling.

\subsection{High-\texorpdfstring{$z$}{z} LAEs as a probe of neutral fraction of the IGM}
\label{ssec:res_lae_xhi}

\begin{table*}
  \centering
    \begin{tabular}{cccc}
    \hline\hline
    Literature & red peak shape & $\vred$ & $\sigred$ \\ \hline
    \citet{Kakiichi16} & Gaussian & 600 km s$^{-1}$ & $20.4 h^{1/3} ( \frac{M_h}{10^8 \Msun})^{1/3} (\frac{1+z}{7.6})^{1/2} \km \s^{-1}$ \\
    \citet{Mason18_baye_igm}  & Gaussian & empirical relation between $\vred$ and halo mass & $\vred$  \\
    \citet{Weinberger_L19} & Gaussian & ${1.5, 1.8} v_{\rm circ}$ & 88 km s$^{-1}$ \\
    \citet{Gangolli21} & Gaussian & $\beta v_{\rm circ}$ & $v_{\rm virc}/2.355$ \\
    \citet{Bruton23_cosmic_var_xhi} & Gaussian & $-3156_{-895}^{+882}+80_{-21}^{+21} \log_{10}\left(L_{\rm Ly\alpha}\right)$ & $\vred$ \\
    \hline\hline
    \end{tabular}
  \caption{Summary of assumptions used for the inference of the IGM neutral fraction $\xhi$ for several observational works. We note that \citet{Mason18_boosted_lya_trans}, \citet{Hoag19}, \citet{Whitler20}, \citet{Bolan22} and \citet{Morishita23} also use the \citet{Mason18_baye_igm} model. }
  \label{tab:lae_xhi}
\end{table*}

\begin{figure}
    \centering
    \includegraphics[width=1.\linewidth]{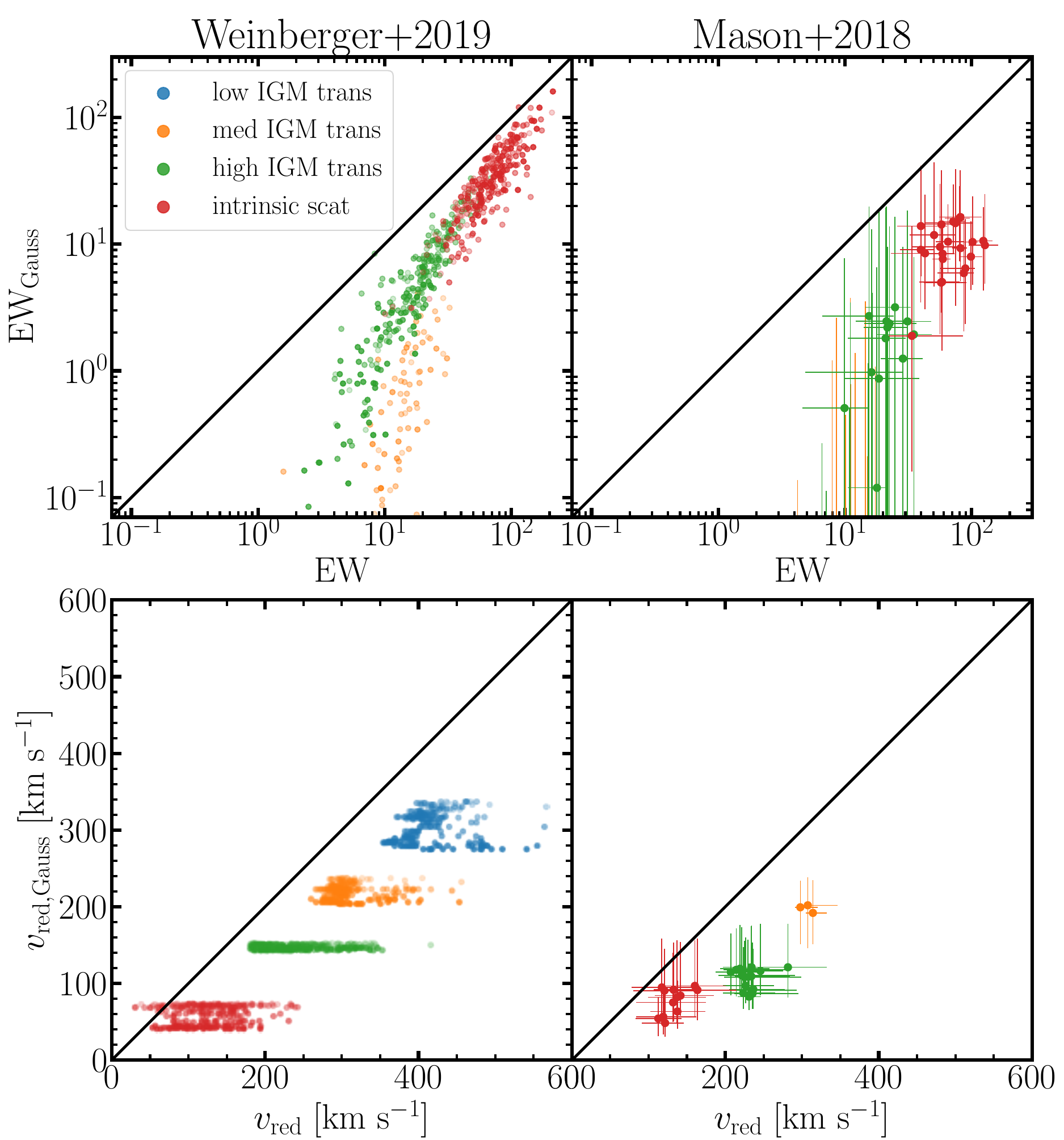}
    \caption{Testing assumptions of intrinsic scattered Gaussian red peaks used in observational work. We compare EW calculated with Gaussian models EW$_{\rm Gauss}$ to the actual EW in the top row, and compare $v_{\rm red, Gauss}$ to $v_{\rm red}$ in the bottom row. The left column shows the results for the \citet{Weinberger_L19} Gaussian model and the right column is for \citet{Mason18_baye_igm}. We note that for the model of \citet{Mason18_baye_igm} we see very few data points of the ``low IGM transmission'' group as either the EW$_{\rm Gauss}$ are too small or the \lya spectra are severely attenuated.
    }
    \label{fig:test_gau_ass}
\end{figure}

The EW PDF of high $z$ LAE has been established as an important probe of the neutral hydrogen fraction of the IGM \citep{Kakiichi16, Mason18_baye_igm, Mason18_boosted_lya_trans, Hoag19, Weinberger_L19, Whitler20, Gangolli21, Bolan22, Bruton23_cosmic_var_xhi, Morishita23}. However, these works usually make several simplifying assumptions about the intrinsic scattered Ly$\alpha$ spectra.

The spectrum of the intrinsic scattered Ly$\alpha$ emission is normally approximated as a simple Gaussian with peak at $\vred$ and width $\sigred$. Different works have made different assumptions about the parameters $\vred$ and $\sigred$, and we list some of them in \autoref{tab:lae_xhi}. In \autoref{fig:test_gau_ass}, we test the assumptions of \citet{Weinberger_L19} and \citet{Mason18_baye_igm}. We see that both of these two Gaussian models underpredict EW and $\vred$, especially at low IGM transmission levels. This is because our simulated LAEs have asymmetric red peaks with $A_f \sim 2.5$. These asymmetric profiles have much broader red wings extending to high velocities of $\sim 400$~km~s$^{-1}$ and beyond. Even with a relatively low value of $\vred$ and $\sigred$ the emission in the red wing is then still able to pass through the IGM and be visible to us. This effect is particularly relevant at low IGM transmissions, with the wavelength range with zero transmission extending to velocities of $\sim 400$~km~s$^{-1}$. For this reason, we recommend assuming asymmetric Gaussian profiles for the red peaks (like the profile in \citealt{Shibuya14_lae_kin_dist})  or adopting more physically motivated simulated Ly$\alpha$ spectra such as those presented here and in \citet{Behrens14, Garel24, Li_Zhihui24, Smith_Aaron25} or better templates from observations at lower redshift with sufficient  S/N to characterise the spectral shape including the extended red wings of the unattenuated \lya emission.

\section{Discussion \& Caveats}
\label{sec:disc}

\subsection{The impact of non-thermal physical processes on \lya and LyC observables }

By investigating how the properties of galaxy evolution and its \lya observables respond across a range of galaxy formation models that incorporate different physical processes, we can systematically disentangle the specific impact of each physical mechanism. By isolating the role of individual processes, we can obtain a clearer framework to interpret observational signatures.

The inclusion of CRs drives the largest differences in the \lya observables. The resulting powerful multi-phase outflows drive low-density channels that facilitate enhanced \lya and LyC radiation escape. While our finding of extended red wings is observed across all models, this result is particularly prominent in the models with CRs, further enhanced when CR streaming and its associated heating is accounted for. This finding is arguably expected, as streaming increases gas-CR coupling and drives more pronounced galactic winds \citep[e.g.,][]{Ruszkowski17, Wiener17} resulting in  additional scattering in the red tail of high velocities. This is reflected in denser outflows ejected with slower velocities than in the no-streaming case. These effects reduce the peak velocity $\vred$ from approximately $200\,\km\,\s^{-1}$ to $\sim100\,\km\,\s^{-1}$, and shift the distribution of measured EWs to higher values, not only displacing the average of the distribution, but also increasing its high velocity tail. Despite these effects, no notable differences are apparent in the surface brightness maps of the \lya emission between the streaming and no-streaming cases.

Across the measured \lya observables, accounting for magnetic fields has a non-negligible albeit secondary impact. One of the mechanisms through which magnetic fields modify our results is their effect on the star formation criterion in the simulations, which requires gas cells to reach higher densities to overcome magnetic support and form stars {--} with a larger number of visible intrinsic \lya clumps in the \lya emission maps, and a somewhat more episodic star formation history. The inclusion of magnetic fields also leads to slightly more extended red wings and EW distributions extending to somewhat larger values at higher redshifts. These findings highlight the necessity to investigate the role of these well known non-thermal physical processes when studying the observable properties of high-redshift LAEs.

We note that the structure, properties and efficiency of CR-driven outflows depend on the CR configuration (such as the CR diffusion coefficients and simulation setup; e.g. \citealt{Dashyan20, Farcy22}). Using tall box ISM simulations, \citet{Gronke18}, \citet{Rathjen23}, and \citet{Armillotta24} find CRs to drive slower, smoother, and cooler outflows that inhibit \lya escape at line center along the direction perpendicular to the disk \citep{Gronke18}. The setup of their simulations allows higher resolutions and they can be seen as as a controlled conditions experiments. However, these simulations do not account for  the cosmological environment and its evolution, they require periodic or shearing-periodic boundary conditions, and they are not able  to capture large-scale structures and the resulting complex flows and asymmetries present in cosmological galaxy formation simulations, both due to  secular evolution and the mergers and interactions that are particularly frequent during Cosmic Dawn. Conversely, our cosmological zoom-in simulations exhibit more bursty star formation, irregular gas flows, and three-dimensional outflow geometries. These setup differences likely explain the formation of low-density escape channels in our simulations, emerging in conditions and geometric configurations not covered by the tall box simulations and leading to our measured high $\fesclya$. A better 
understanding of the interplay between small-scale CR wind physics and large-scale galaxy structure remains an important direction for future work.

\subsection{Double-peaked and anomalously bright high-z LAEs}

There is a number of rather puzzling high-$z$ LAEs that our simulations cannot explain.

Surprisingly, a number of double-peaked Ly$\alpha$ spectra have been observed, such as COLA1 \citep{Hu_EM16, Matthee18, Torralba-Torregrosa24}, NEPLA4 \citep{Songaila18}, A370p\_z1 \citep{Meyer21} and CDFS-1 \citep{Moya-Sierralta24}. Such profiles had been thought to be highly unlikely at high $z$ and are hard to reproduce in simulations. Indeed, our simulation does not predict any observed double-peaked \lya spectra (c.f. \autoref{fig:lya_spec_diff_dust} in \aref{asec:model_dw_inst}). Observation of prominent blue peaks may imply that reionization is driven partially by sources more luminous than star forming galaxies, such as faint AGN \citep{Padmanabhan&Loeb21, Asthana24_faint_agn_reion, Protusova24}, producing ionised bubbles with very low residual neutral fraction in their vicinities aiding the transmission of the blue peaks. Another possible explanation is that these double-peaked LAEs are associated with overdense environments of star-forming galaxies driving rapidly large ionization bubbles facilitating the escape of the blue component \citep{Torralba-Torregrosa24, Songaila24}. We lastly note that interpreting these objects may need better  modelling  of the  blue peaks of the \lya emission \citep{Gurung-Lopez19_flareon, Hutter23, Garel24}.

Recently, \citet{Witstok24_lae_z13} discovered a very strong LAE JADES-GS-z13-1-LA at an extremely high redshift of $z \sim 13$. This LAE has a surprising large observed value EW$=43_{-11}^{+15}\angstrom$ and inferred intrinsic scattered value EW$=686_{-249}^{+461}\angstrom$
\footnote{As the systemic redshift of this object is unknown, this value is very sensitive to the assumptions of the intrinsic scattered line profile. For the details of the modelling see \citet{Witstok24_lae_z13}.}, 
significantly larger than the typical values of our simulated LAEs (c.f. \autoref{fig:lya_time_evo_diff_phys}). Due to the high observed EW, its red wing likely has a large velocity offset, estimated to be $\vred \sim 700\pm 300$~km~s$^{-1}$, extending further to the red than the most extreme spectra in the RTCRiMHD simulation even without dust (c.f. \autoref{fig:deci_fesc_lya}), but note the rather large error of the observed  $\vred$. Modelling the \lya emission of such strong LAEs at this early redshift is clearly a theoretical challenge and we leave this for future work.

\subsection{Other caveats}
\label{ssec:caveats}

In this section, we first discuss the caveats associated with our physical model of galaxy formation and then of our treatment of the synthetic observations.

The \textsc{Azahar} simulation suite builds on the pathfinder  \textsc{Pandora} project and includes the same physical processes. A number of caveats regarding the thermodynamics and feedback implementations have already been discussed in \citetalias{Yuan_Yuxuan24} and \citet{Martin-Alvarez23}. Here, we further detail potential caveats regarding our galaxy formation and feedback model. 

An important caveat when employing this type of numerical model to investigate \lya emission is the spatial resolution of the simulations. Specifically, quantities such as the LyC escape fraction and star formation may only reach convergence at resolutions of better than 1~pc \citep{Kimm17}. No fully cosmological or cosmological zoom-in simulations to date \citep{A_Smith19, Garel21, A_Smith22a, A_Smith22b, Blaizot23, Choustikov24} reach such resolutions, despite all being at the limit of what is currently computationally feasible. This resolution limitation is likely to have  notable effects, and  it may e.g. lead to excess blue wing emission \citep{Hayes21}. Note further that \citet{McCourt18} found cool atomic gas to have a misty structure composed of tiny cloudlets on even smaller, sub-pc scales. Clumpy structure on such small scales cannot be resolved in our simulations, but will affect the \lya transport below the resolution scale. This suggests the need for subgrid modelling to mitigate the effect of insufficient resolution on the \lya radiative transfer, and will require future work. Based on previous studies, we generally expect more significant \lya escape \citep{Laursen13} and a directional boost of EWs \citep{Gronke&Dijkstra14} in more clumpy media.

\textsc{Azahar} adopts a standard IMF, while several recent studies find evidence of a top-heavy IMF at high redshift \citep{Bromm09, Bromm13, Susa14}. \citet{Sharda22} have shown that low metallicities of $Z\sim 10^{-6}\,{\mathrm Z_\odot}$ at high redshift can lead to a top-heavy IMF with a characteristic mass of $\sim 50\,\Msun$ while \citet{Sharda23} have illustrated that deviations of carbon and oxygen abundances from solar scalings at low metallicity increase the characteristic mass of the IMF. \citet{Cameron24} and \citet{Katz24} have proposed that the strong Balmer jump and the two-photon continuum from nebular emission are possible indicators of a top-heavy IMF at high $z$. A top-heavy IMF will increase the fraction of massive stars, thus boosting the efficiency of SNe feedback, creating more low-density channels for \lya photons to escape. The enhanced stellar feedback effects may increase the outflow velocity, leading to redder line profiles. It also affects the ionization structure of the ISM, possibly enhancing the production of the intrinsic \lya emission.

Recently, there have been several notable theoretical efforts in the modelling of galactic outflows. An important frontier is multiphase outflows. \citet{Rey24} have pointed out that resolving the cooling length of the outflows significantly increases their mass and energy loading factor. \citet{Yuan_Yuxuan23}, \citet{Fielding22} and \citet{Tan&Fielding24} have discussed why the evolution of cool clouds in outflows is crucial for the study of their multiphase nature. Other advances include the inclusion of \lya radiation pressure. \citet{Kimm18}, \citet{Tomaselli&Ferrara21} and \citet{Nebrin25} have suggested that \lya radiation pressure exerted by luminous objects can dominate over direct radiation pressure, especially at cosmic dawn. Other important advances have been made in the theoretical modelling of CR. \citet{Armillotta21, Armillotta22, Armillotta24} have shown that physically modelled CR diffusion coefficients can vary up to 4 orders of magnitude depending on the local ISM conditions. \citet{Sampson23} have argued that CR diffusion on ISM scales shows super-diffusion behaviour, effectively giving different ``classical'' diffusion coefficients at different spatial scales. Several works also begin to incorporate spectrally resolved CR \citep{Girichidis20, Girichidis22, Hopkins23a_pl_cr}. Accounting for the processes described above will typically increase the total outflowing mass budget, potentially reinforcing the red wing signature. However, configuration variations such as higher CR diffusion coefficients will reduce the proportion of entrained cold and neutral gas, increasing escape fractions and reducing the red peak velocities.

This first iteration of the \textsc{Azahar} simulations does not include AGN feedback, which may play a role in shaping the largest galaxies at high redshifts \citep{Kumari24, Hegde24, Bennett24, Juodzbalis24}. There is also now more evidence of some \lya emission driven by QSOs \citep{Farina19} and AGN-driven jets \citep{Wang_Wuji23}, which we will address with the second generation of \textsc{Azahar} models, including supermassive black holes. Another interesting question is the connection between AGN orientation angle and the morphologies of giant \lya nebulae, which we leave for future work.

We lastly discuss the impact of dust physics as the dust attenuation could have a comparable effect to the IGM attenuation \citep{Xu_C23}. Several works have suggested variation of dust attenuation curves at high $z$ \citep{Shen_X20, Mushtaq23, Kumari24, Sanders24}. Dust scaling relations are still very unconstrained both in observations and theory. Several works have shown that the dust content is not a linear function of metallicity as we assumed in this work \citep{Remy-Ruyer14, McKinnon17}. Furthermore, \citet{Nelson18, Qiu19, Vogelsberger20, Cullen23, Caputi24} and others have suggested that the dust content decreases at high $z$. As we have seen earlier, \lya emission is very sensitive to dust even at low $A_V$, due to the resonant scattering. In future, we plan to extend the \textsc{Azahar} simulation suite with simulations employing the self-consistent and physically motivated dust model of \citet{Dubois24}, to further explore this problem. Note, however, that dust may not have enough time to grow at high redshift, leading to a reduced dust content compared to the fiducial case of this work. As already discussed, a reduced dust content predicts more extended red wings and higher \lya escape.

We also emphasize that our modelling of the synthetic observations is still quite simplistic. For example, due to the finite size of the aperture and the surface brightness detection limit, the instruments will not capture the diffuse Ly$\alpha$ emission in the outskirts of galaxies. \citet{Bhagwat24} have suggested that the spatial offset between UV and Ly$\alpha$ may cause slit-losses of $\approx 65\%$ and that misplaced slits can result in significantly underestimated \lya EWs and escape fractions.

\section{Conclusions}
\label{sec:con}

In this work we leverage a new suite of cosmological radiative transfer simulations to investigate the spatial and spectral distribution of \lya emission along the merger sequence of a massive disc galaxy progenitor to study the visibility of reionization-epoch LAEs. The employed models are a subset of simulations from the state-of-the-art \textsc{Azahar} suite that self-consistently model the frequent mergers of disc progenitors at high $z$ with a range of different physical processes included. We have post-processed our simulations with the Monte-Carlo radiative transfer code \textsc{Rascas} and convolved the output spectra with published IGM damping wing curves extracted from the \textsc{Sherwood-Relics} reionization simulation suite. We have further convolved the spectra with the JWST LSF to obtain realistic synthetic LAE spectra. We have compared our results with the JADES and other high-$z$ observations of LAEs and have tested several assumptions made when inferring the neutral fraction of  IGM $x_{\rm H~\textsc{i}}$ from the visibility of reionization-epoch LAEs. We summarize our main findings below.

\begin{itemize}
    \item \textit{Galaxy merger sequence}: Galaxy merger and feedback cycles drive the evolution of SFR, dust enrichment, and galaxy morphology. These then strongly modulate the intrinsic scattered \lya and LyC observables, all with large variations between different LOS \citep{Barnes10}. Different simulated physics also have a large impact on the spatial distribution and spectral properties of the \lya emission during the merger sequence. The simulations including cosmic rays show more bursty star formation histories, lower dust content, lower $\vred$ and higher escape of Ly$\alpha$/LyC photons.
    \item \textit{Red peak of \lya spectra}: All of our simulated LAEs' intrinsic scattered spectra show strongly asymmetric red peaks which can extend beyond $400$~km~s$^{-1}$. These extended red wings aid transmission through the IGM leading to significantly larger \lya EWs compared to Gaussian line profiles.
    \item \textit{Effects of galaxy formation physical modelling variations:} We find that the inclusion of cosmic ray feedback results in the largest differences across our simulations. Cosmic ray feedback leads to more asymmetric \lya emission with extended red wings, an increase of $\fesclya$ of about 1~dex, and a more prominent tail of lines of sights towards higher EW. Both cosmic ray streaming and magnetic fields drive smaller but notable secondary effects. The former increase the red wing asymmetry throughout the simulation. Magnetic fields have a more subtle impact {--} slightly increasing outflow velocities at high redshifts. 
    \item \textit{Comparison with observations}:  Variations of the \lya emission along the merger sequence and along different lines-of-sight of our single galaxy simulated with cosmic ray-driven outflows remarkably reproduce the $\vred$-EW, $\fesclya$-$\vred$, $\fesclya$-EW, and $\fesclya$-$\LLya$ relations observed in the \textsc{Muse} and \textsc{Jades} surveys if we assume that there is little or no dust and some moderate slit losses in the JWST observations. We further find that IGM attenuation is the main driver of these relations.
    \item \textit{LAE-based inference of neutral fraction}: We test the assumptions of Gaussian red peaks commonly adopted. We find that this significantly underestimates EW and $\vred$, with increasing discrepancy for decreasing IGM transmission levels.
    \item \textit{Dependence of visibility on dust and the modelling of galactic outflows}: The visibility of LAEs during the EoR is particularly sensitive to the modelling of dust as well as to the physical processes responsible for the occurrence of galactic outflows.
\end{itemize}

Overall, we confirm the finding of observing galaxy mergers along fortunate LOS as the frequent origin of high-$z$ LAEs, advocated by \citetalias{Witten24}. We emphasize the necessity to use physically motivated asymmetric \lya spectra or good observational templates for accurately constraining the reionization history. Despite all these complexities, with JWST pushing LAE observations to higher and higher $z$, interpretations of this large LAE population taking into account the orientation effects and merger sequence evolution should ultimately provide valuable clues for how and when the Universe was reionized as well as for the evolution of galaxies in the reionization era.

\section*{Acknowledgements}
YY thanks for the discussions with Gareth Jones, Charlotte Simmonds and Callum Witten. YY is supported by Isaac Newton's studentship from the Cambridge Trust. 
SMA acknowledges support by a Kavli Institute for Particle Astrophysics and Cosmology (KIPAC) Fellowship. SMA is also supported by the NASA/DLR Stratospheric Observatory for Infrared Astronomy (SOFIA) under the 08\_0012 Program. SOFIA is jointly operated by the Universities Space Research Association,Inc.(USRA), under NASA contract NNA17BF53C, and the Deutsches SOFIA Institut (DSI) under DLR contract 50OK0901 to the University of Stuttgart. SMA is also supported by the NASA Astrophysics Decadal Survey Precursor Science (ADSPS) Program (NNH22ZDA001N-ADSPS) with ID 22-ADSPS22-0009 and agreement number 80NSSC23K1585. MGH and DS have been supported by STFC consolidated grants ST/N000927/1 and ST/S000623/1. 
This research was supported in part by grant NSF PHY-2309135 to the Kavli Institute for Theoretical Physics (KITP). Part of the work has been performed as part of the DAE-STFC collaboration `Building Indo-UK collaborations towards the Square Kilometre Array' (STFC grant reference ST/Y004191/1). JW gratefully acknowledges support from the Cosmic Dawn Center through the DAWN Fellowship. The Cosmic Dawn Center (DAWN) is funded by the Danish National Research Foundation under grant No. 140.
Both the \textsc{Azahar} simulation suite and the \textsc{Rascas} radiative transfer simulations in this work were performed on the DiRAC@Durham facility managed by the Institute for Computational Cosmology on behalf of the STFC DiRAC HPC Facility (www.dirac.ac.uk). The equipment was funded by BEIS capital funding via STFC capital grants ST/P002293/1, ST/R002371/1 and ST/S002502/1, Durham University and STFC operations grant ST/R000832/1. DiRAC is part of the National e-Infrastructure. This work was performed using resources provided by the Cambridge Service for Data Driven Discovery (CSD3) operated by the University of Cambridge Research Computing Service (www.csd3.cam.ac.uk), provided by Dell EMC and Intel using Tier-2 funding from the Engineering and Physical Sciences Research Council (capital grant EP/P020259/1), and DiRAC funding from the Science and Technology Facilities Council (www.dirac.ac.uk).

\section*{Data Availability}

The data underlying this article will be shared on reasonable request to the corresponding author.



\bibliographystyle{mnras}
\bibliography{all}



\appendix
\section{Further study on the impact of dust}

\subsection{The impact of dust on Ly\texorpdfstring{$\alpha$}{alpha} emission maps of the fiducial RTCRiMHD simulation. }

\begin{figure*}
    \centering
    \includegraphics[width=1.\linewidth]{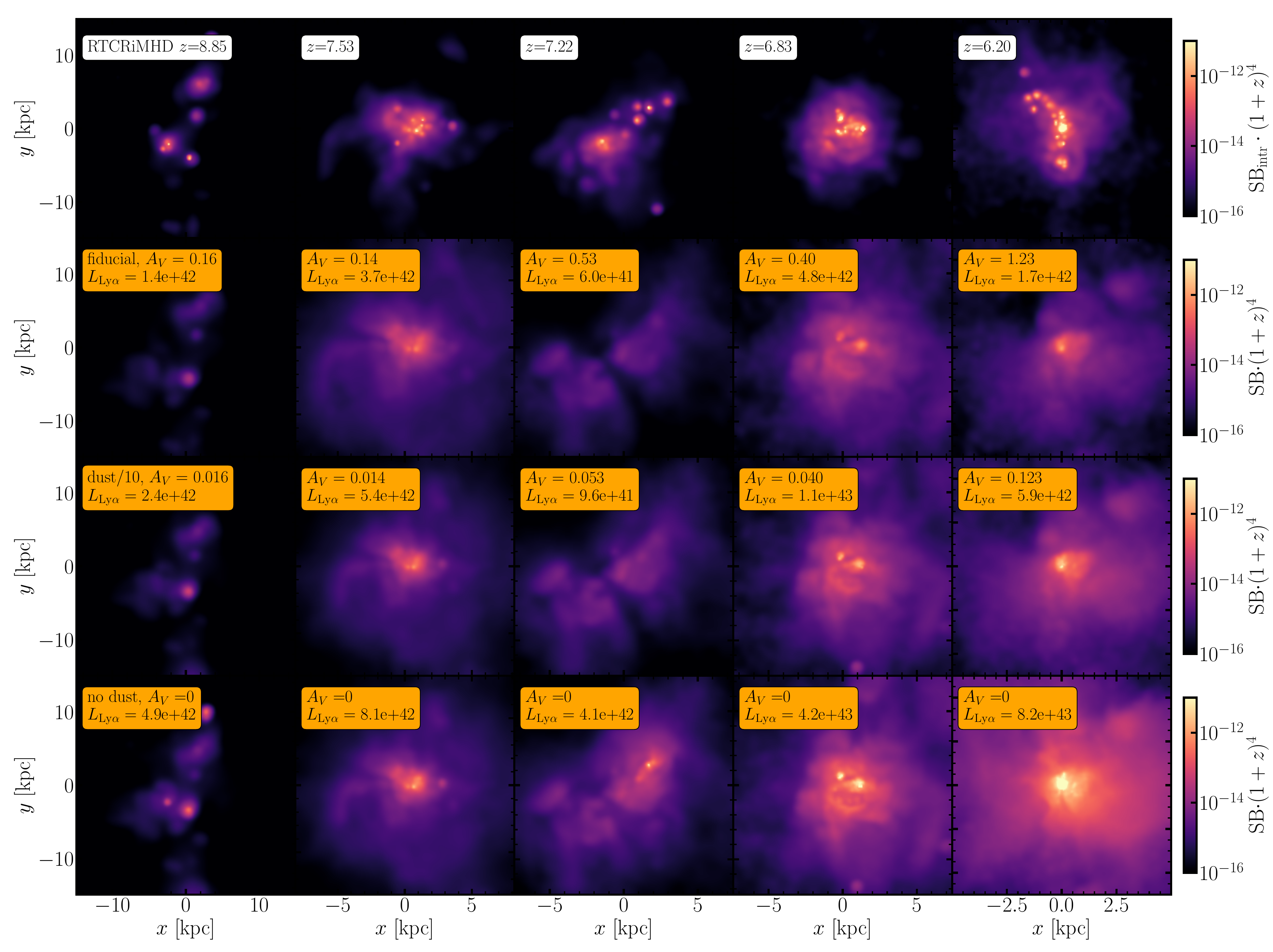}
    \caption{The impact of dust content on Ly$\alpha$ emission across the merger sequence for the  RTCRiMHD simulation. The first row shows maps of the intrinsic Ly$\alpha$ emission while the second to fourth rows show the intrinsic scattered emission for the fiducial dust content, dust content divided by 10, and no dust content, respectively. In the first row, we mark the simulation name and redshifts on the top left of each panel. From the second to fourth rows, we show the dust model with the values of $A_V$  and \lya luminosity on the top left of each panel. The unit of $L_{\rm Ly\alpha}$ is erg s$^{-1}$.
    }
    \label{fig:lya_img_diff_dust}
\end{figure*}

As described in the main text \lya emission is very sensitive to the presence of dust. To illustrate this further, in \autoref{fig:lya_img_diff_dust} we compare maps of the intrinsic \lya emission and the \lya emission scattered by the ISM/CGM for our fiducial RTCRiMHD simulation with its fiducial dust model as well as with the dust reduced by a factor of ten and no dust. The strong effect of even very small amounts of dust corresponding to dust attenuation magnitudes $A_V\sim 0.1$ is clearly visible and can reduce \lya luminosities by factors of 2-50. 

\subsection{The impact of $f_{\rm ion}$ on our results}

\begin{figure*}
    \centering
    \includegraphics[width=.9\linewidth]{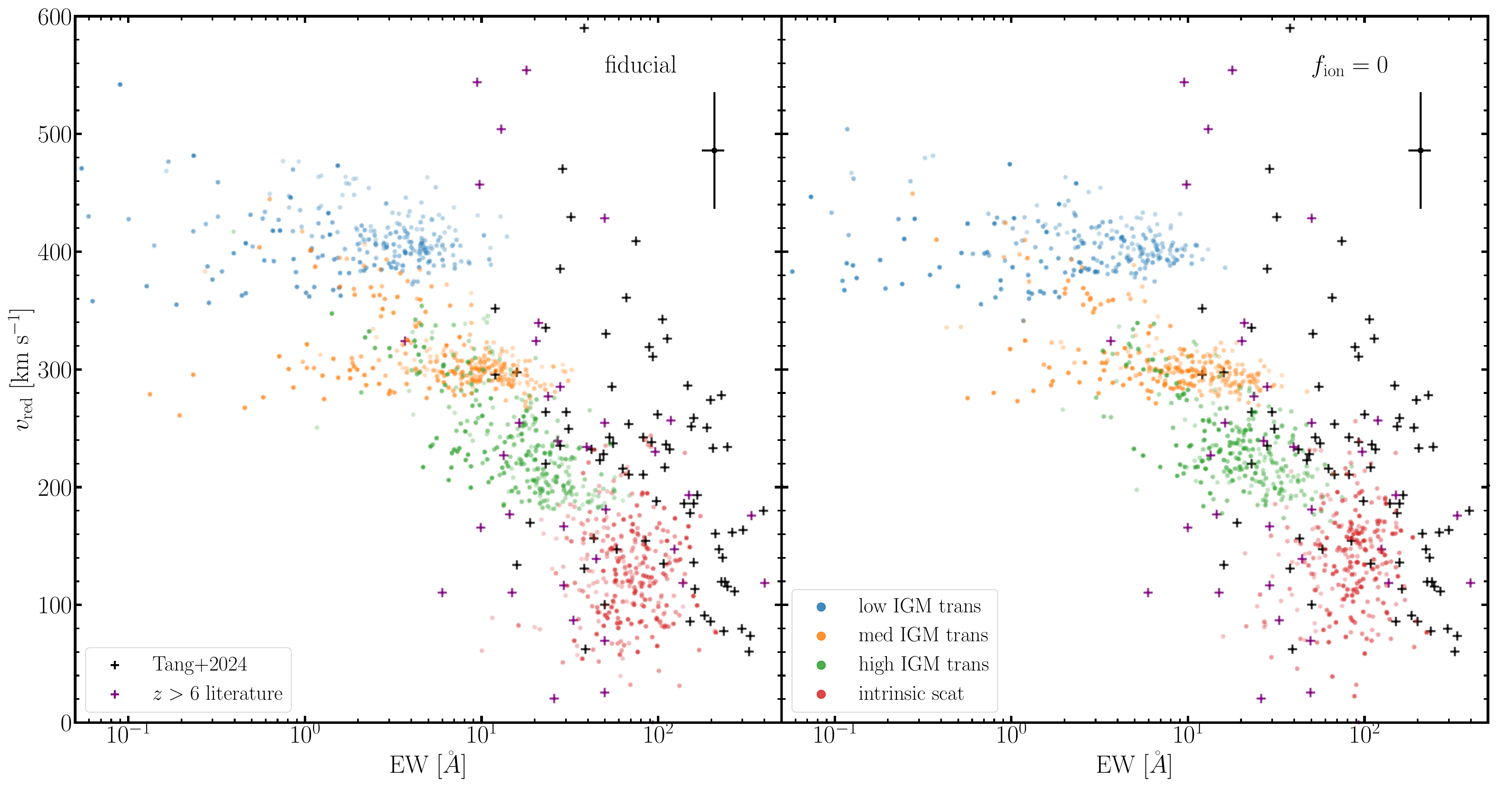}
    \caption{$\vred$ versus EW, but showing variations to the dust model parameter $f_{\rm ion}$  as indicated by the legend ($Z_0=0.005$ for both cases), for the RTCRiMHD simulation. Different colours represent different IGM transmission levels. Higher opacities of markers represent higher redshift. We overplot the same set of observational data \citep{Tang_Mengtao24_lya_z5_6} as in the left panel of \autoref{fig:vred_ew}. Changing $f_{\rm ion}$ causes little change to our simulated results, suggesting that the dust content in ionized gas is low.}
    \label{fig:vred_ew_fion}
\end{figure*}

We show the impact of $f_{\rm ion}$ on the $\vred$ - EW relation in \autoref{fig:vred_ew_fion}. We see that there are little variations of results when $f_{\rm ion}$ goes from $0.01$ to $0$. This means that the \citet{Laursen09_lya_dust} dust model predicts low dust content in ionized gas.

\section{More detailed comparison of the physical properties governing the Ly\texorpdfstring{$\alpha$}{alpha} emission during the merger sequence in the four different simulation runs.}
\label{asec:add_anal_merger}

\begin{figure*}
    \centering
    \includegraphics[width=1.\linewidth]{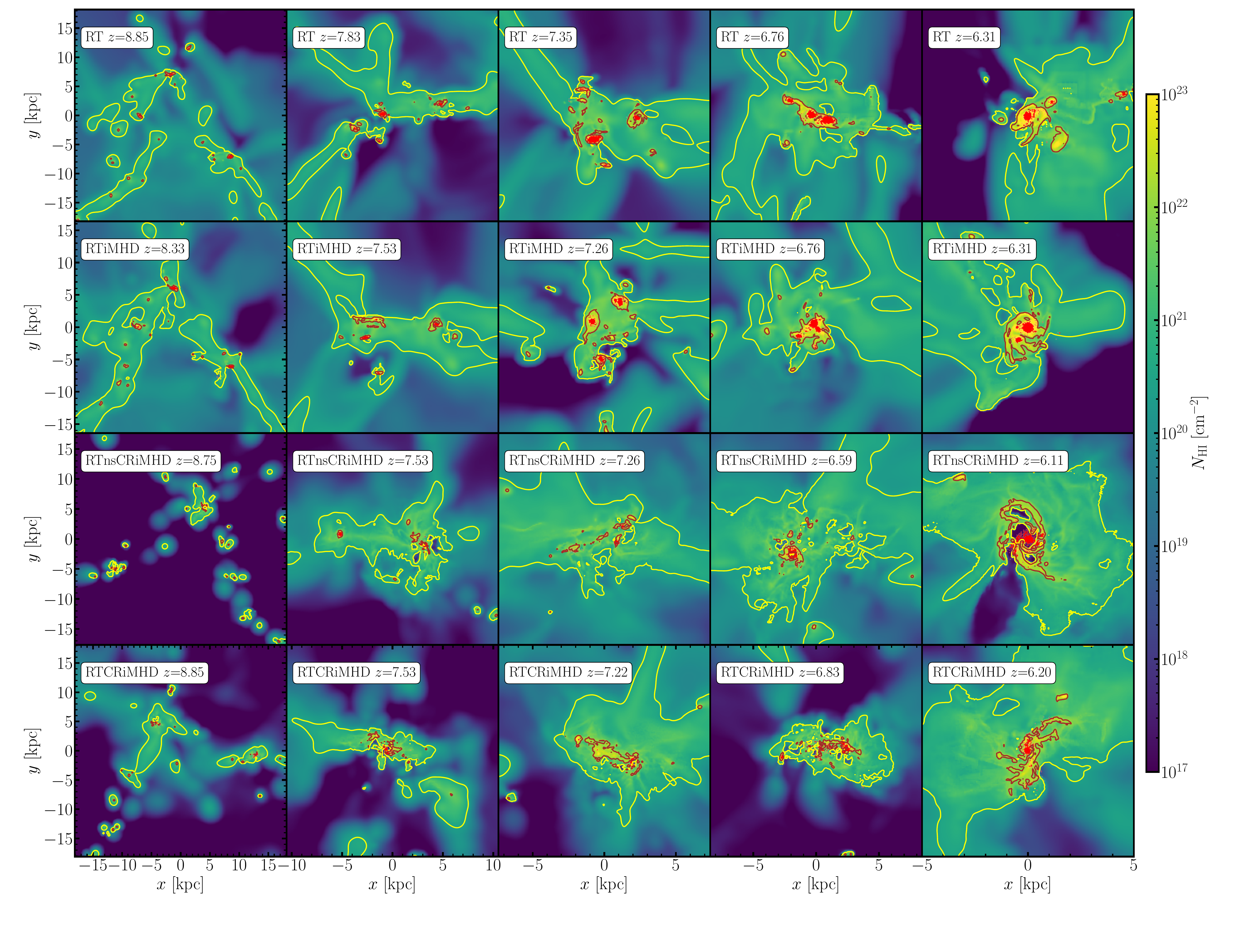}
    \caption{Maps of neutral hydrogen column density $\NHI$ across the merger sequence for simulation runs with different physics included. Different rows show the results of different simulation runs. The yellow (brown) contours represent column densities of $N_{\rm H~\textsc{i}} = 3\times 10^{20} \mathrm{cm}^{-2}$ ($10^{22} \mathrm{cm}^{-2}$). The red dots represent young star particles with age $t_*<30$~Myr.}
    \label{fig:img_diff_phys_nhi}
\end{figure*}

\begin{figure*}
    \centering
    \includegraphics[width=1.\linewidth]{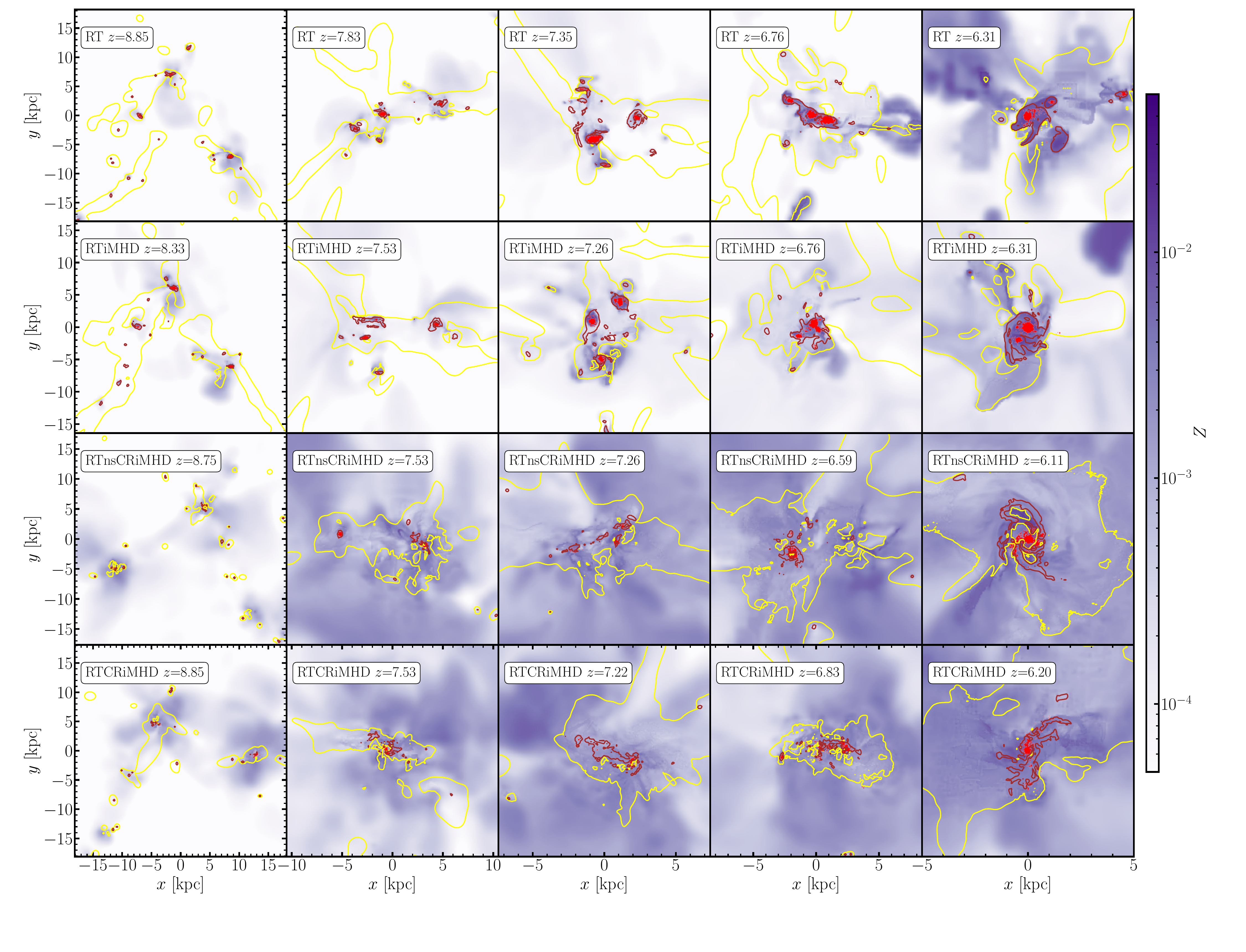}
    \caption{Same as \autoref{fig:img_diff_phys_nhi}, but showing  projected maps of H~\textsc{i}-mass weighted metallicity $Z$.}
    \label{fig:img_diff_phys_Z}
\end{figure*}

In \autoref{fig:img_diff_phys_nhi} we show maps of H~\textsc{i} column density $\NHI$ along the merger sequence for the simulation with different physics implementations. For the simulations including cosmic rays, the three approaching galaxies are in an environment with much lower gas densities in the isolated phase. In the merger and disc phase, the simulations including cosmic rays show more disturbed morphologies and larger numbers of low density channels due to the more efficient galactic winds. In the CR simulations the young stellar particles are also more likely to be located within the escape channels, resulting in larger $\fesclyc$. We show maps of projected H~\textsc{i}-mass weighted metallicity $Z$ along the merger sequence in \autoref{fig:img_diff_phys_Z}. We clearly see that the strong winds in the CR simulations transport metals out to much larger radii, whereas the metal distributions in the RT and RTiMHD simulations are very compact, efficiently shielding most of the intrinsic \lya emission.

\begin{figure*}
    \centering
    \includegraphics[width=1.\linewidth]{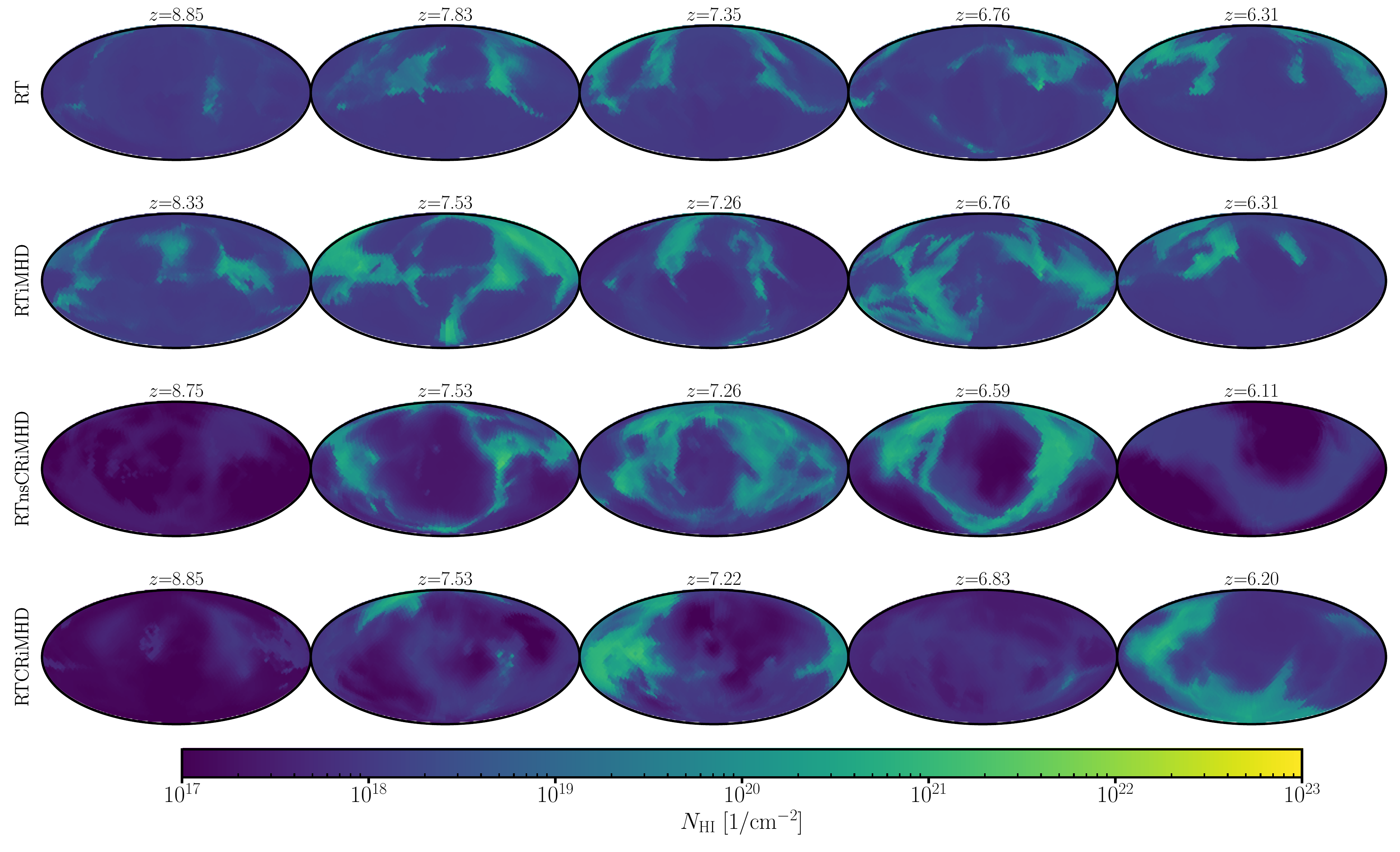}
    \caption{Angular distribution of neutral hydrogen column density $N_{\rm H~\textsc{i}}$, for the same configuration as \autoref{fig:img_diff_phys_nhi}.
    }
    \label{fig:ang_nhi_diff_phys}
\end{figure*}

\begin{figure*}
    \centering
    \includegraphics[width=1.\linewidth]{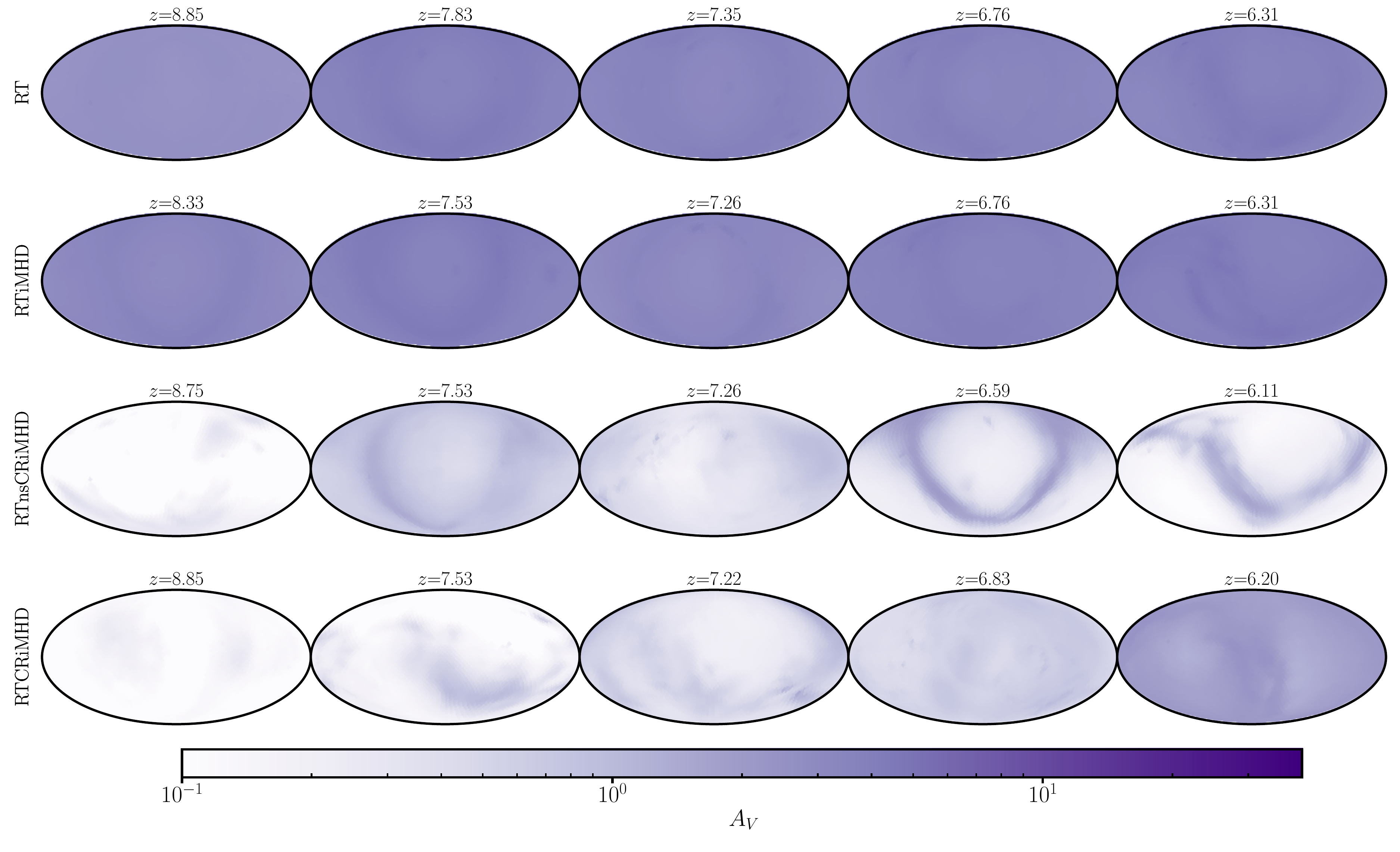}
    \caption{Angular distribution of dust attenuation $A_V$, for the same configuration as \autoref{fig:img_diff_phys_nhi}. }
    \label{fig:ang_av_diff_phys}
\end{figure*}

\begin{figure*}
    \centering
    \includegraphics[width=1.\linewidth]{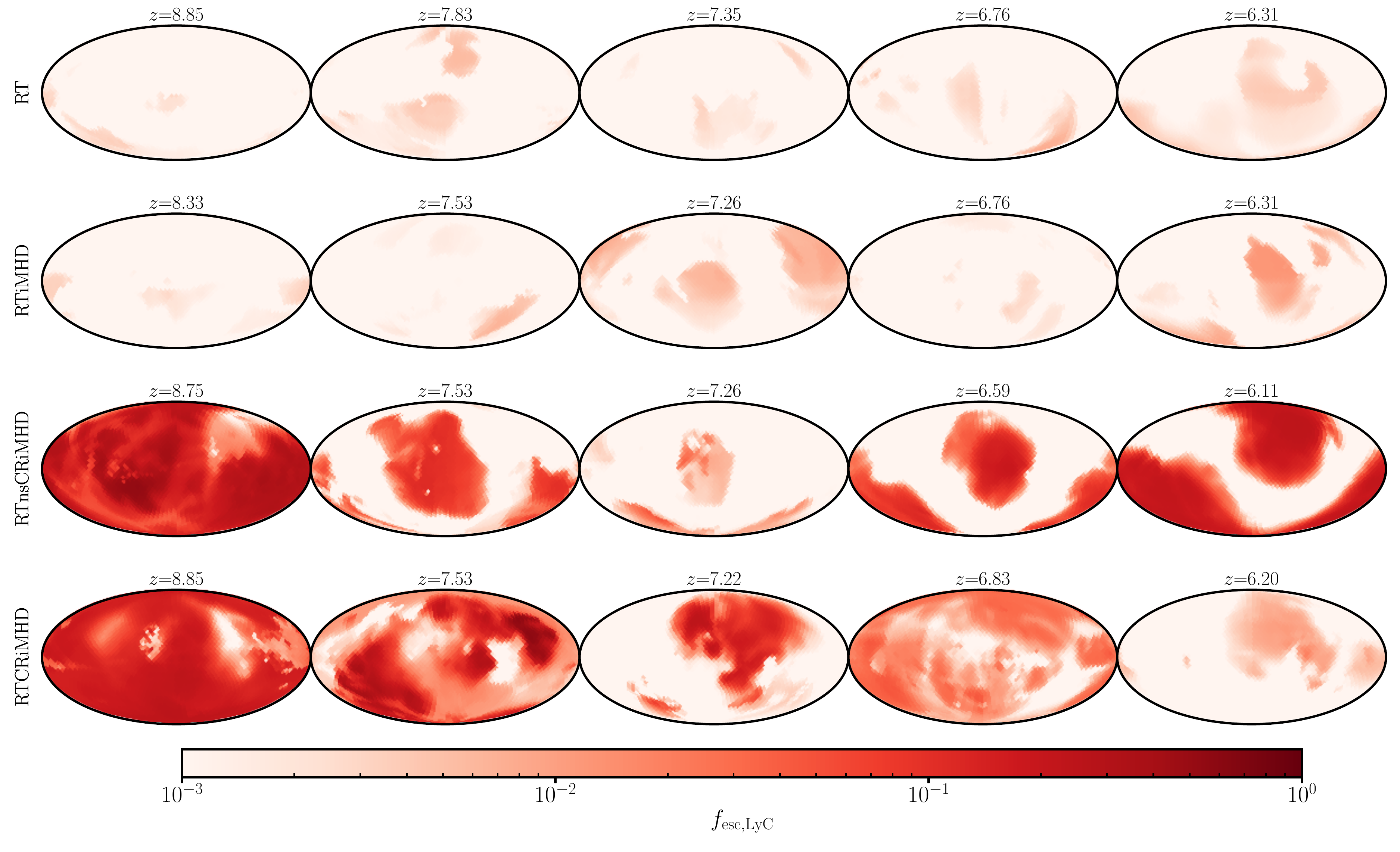}
    \caption{Angular distribution of LyC escpape fraction $\fesclyc$, for the same configuration as \autoref{fig:img_diff_phys_nhi}. 
    }
    \label{fig:ang_fesc_lyc_diff_phys}
\vspace{10cm}
\end{figure*}

We show the angular distributions of $\NHI$, $A_V$ and $\fesclyc$ for the different simulations in \autoref{fig:ang_nhi_diff_phys}, \autoref{fig:ang_av_diff_phys}, and \autoref{fig:ang_fesc_lyc_diff_phys}, respectively. Comparing $\NHI$ and $\fesclyc$ distribution, we see that the stronger outflows in the CR simulations create channels with lower densities, aiding the escape of LyC photons. The $A_V$ distributions are rather isotropic with a high value of $\sim 4$ in the RT and RTiMHD simulations because of the compact distribution of the dust. In comparison, the CR simulations show anisotropic $A_V$ distributions that follow more closely the $\NHI$ distributions, but are significantly smoother. The CR simulations also have significantly lower values of $A_V$.

\section{Modelling of damping wings and JWST LSF}
\label{asec:model_dw_inst}

\begin{figure*}
\begin{subfigure}{0.2186\textwidth}
  \centering
  \includegraphics[width=\textwidth]{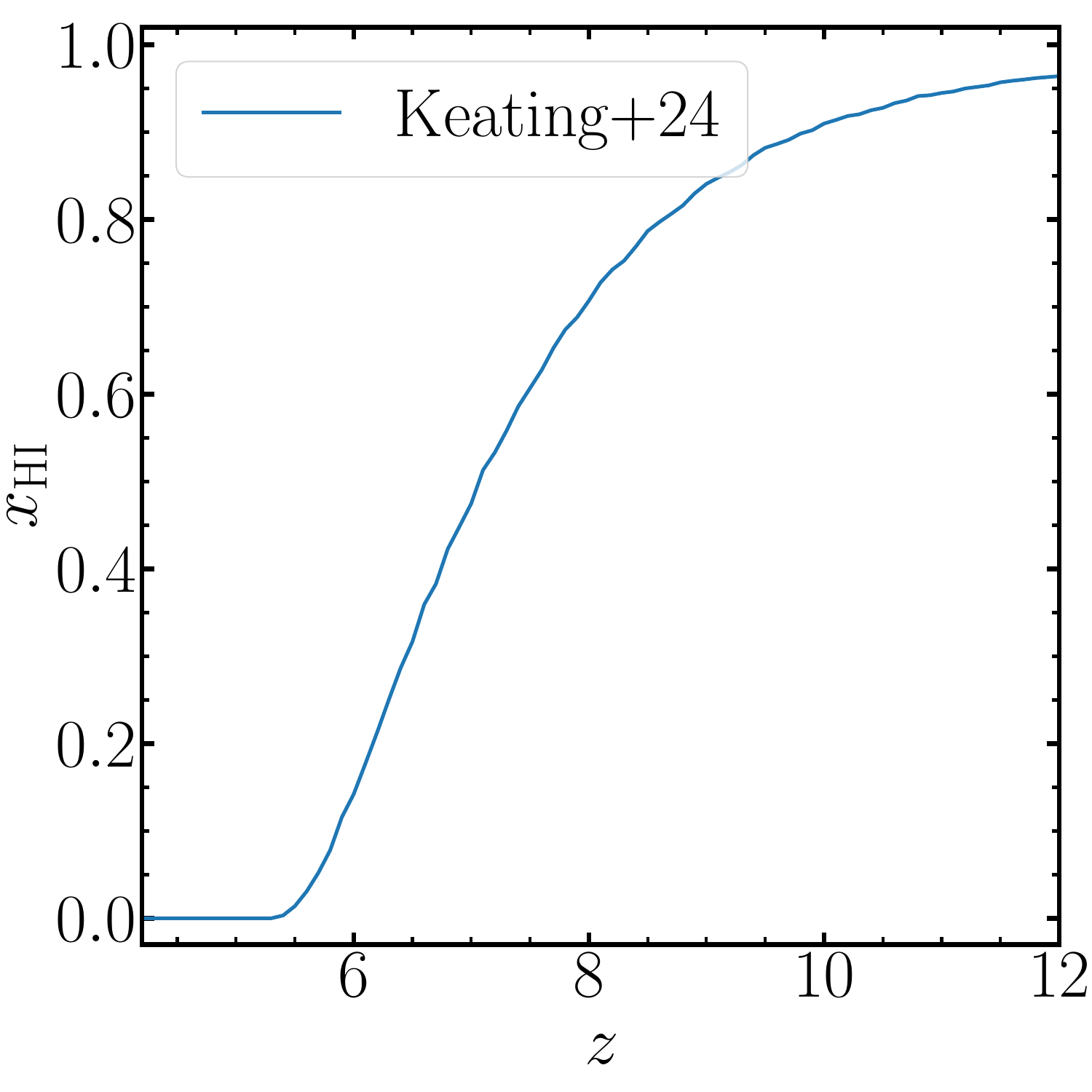}
  \caption{}
\end{subfigure}\hfill
\begin{subfigure}{0.7814\textwidth}
  \centering
  \includegraphics[width=\textwidth]{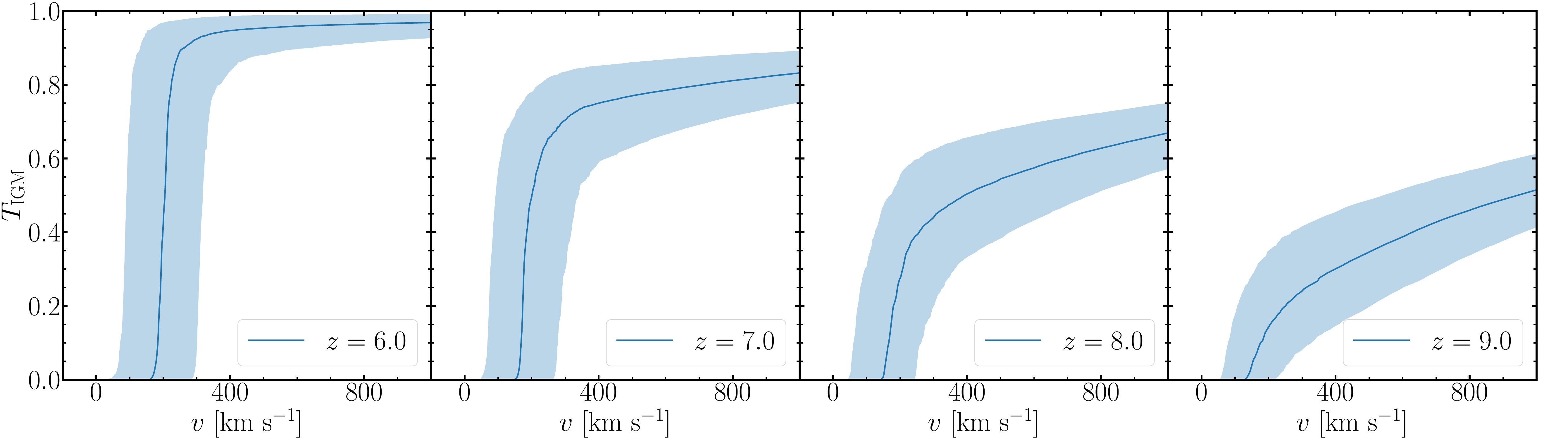}
  \caption{}
\end{subfigure}\hfill
\caption{(a) Reionization history of \citet{Keating24_jwst_dw_cos}. (b) IGM damping wing curves from \citet{Keating24_jwst_dw_cos}. The zero transmission range of the damping wings of different IGM transmission levels extends to different velocities, while at fixed transmission level, the overall transmission fraction decreases at higher redshifts. }
\label{fig:igm_dw}
\end{figure*}

In this section we show the reionization history and the IGM damping wing predictions of  \citet{Keating24_jwst_dw_cos} in \autoref{fig:igm_dw}. We show the evolution of the neutral fraction on the left and the redshift evolution of the IGM attenuation curves with their 1$\sigma$ uncertainties on the right. In the simulation reionization has a midpoint of $z\sim 7.2$ and completes at $z\sim 5.3$. The damping curve has a steep suppression of the transmission due to IGM infall and residual H~\textsc{i} within the ionized bubbles that reach to the red of the systemic redshift. The velocities to which zero transmission extends have a large scatter of $\sim 50-300$~km~s$^{-1}$, depending on the distance of the galaxy from the infalling bubble wall in velocity space. The attenuation is therefore very sensitive to whether the red peak extends beyond $\sim 300$~km~s$^{-1}$. We define the   IGM transmission curves  $1\sigma$ lower and higher than the median as low, and high IGM transmission levels, respectively.

Most of the JWST observations of LAEs \citep{Saxena24, Jung24, Chen_Z24, Witstok25_pri_lae, Jones24_emer_evol_la, Napolitano24} have been conducted in NIRSpec G140M mode\footnote{The reader can refer to this JWST website link \href{https://jwst-docs.stsci.edu/jwst-near-infrared-spectrograph/nirspec-instrumentation/nirspec-dispersers-and-filters\#gsc.tab=0}{NIRSpec Dispersers and Filters} for the resolving power of JWST observing modes.}. 
We have convolved the mock spectra from the simulations with the line spread function (LSF) for the G140M mode, following the procedures described in \citet{Jones24_emer_evol_la}. We adopt a Gaussian LSF with a width $\sigma = \lambda_{\rm Ly\alpha}/(2.355R)$. 

\begin{figure*}
    \centering
    \includegraphics[width=1.\linewidth]{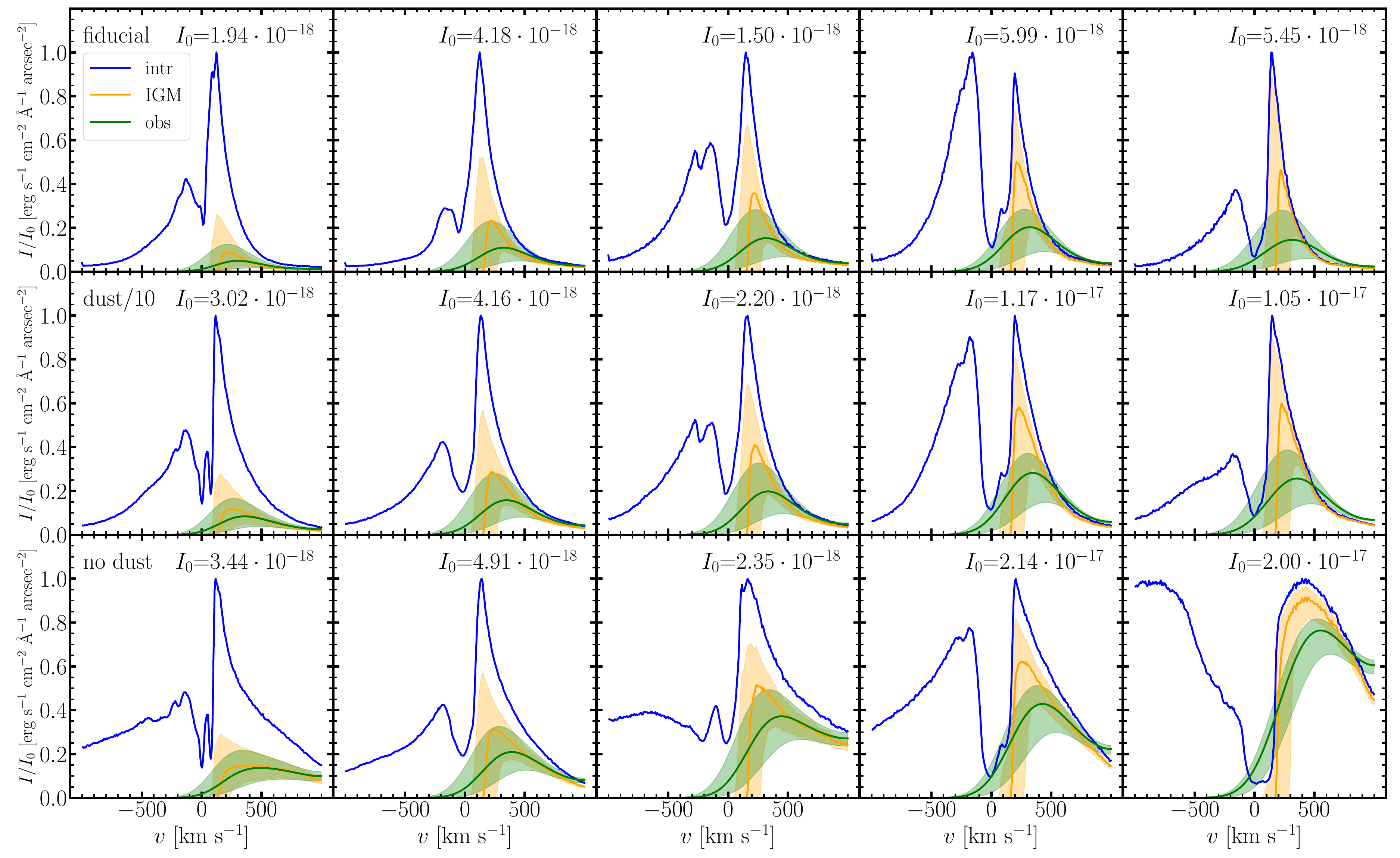}
    \caption{The impact of IGM attenuation and JWST LSF  on the \lya spectra, for the RTCRiMHD simulation. We show the same set of snapshots as in \autoref{fig:lya_img_diff_dust}. From top to bottom we show the results for the fiducial dust content, dust content divided by 10, and no dust content. The blue curves represent intrinsic scattered \lya spectra. The yellow curves show IGM-attenuated spectra for the median IGM transmission curves, while the lower and upper bounds of the yellow shaded region correspond to low and high IGM transmission levels. The green curves with shaded regions show mock observed spectra where we have convolved the IGM-attenuated spectra with the LSF of the JWST G140M mode. 
    }
    \label{fig:lya_spec_diff_dust}
\end{figure*}

We show how IGM damping and JWST LSF affect \lya spectra in \autoref{fig:lya_spec_diff_dust}. We see that the IGM damping wings completely absorb the blue peak of the \lya spectra and higher IGM transmission levels generally predict red peaks with wider profiles and higher flux. The instrumental effects of the JWST G140M mode further smooth the IGM attenuated spectra and the resulting observed spectra are broader and more symmetric, and can even extend to negative velocities on the blue side. Reducing the amount of dust broadens the intrinsic scattered \lya profile and makes it more asymmetric\footnote{In the rightmost column, the intrinsic scattered \lya spectra change significantly between the fiducial dust and the no dust case. This is due to the relatively large dust content in this snapshot of the fiducial RTCRiMHD simulation (c.f. \autoref{fig:ang_av_diff_phys}).}.

\section{Convergence tests of \textsc{Rascas} simulations}

\begin{figure*}
    \centering
    \includegraphics[width=1.\linewidth]{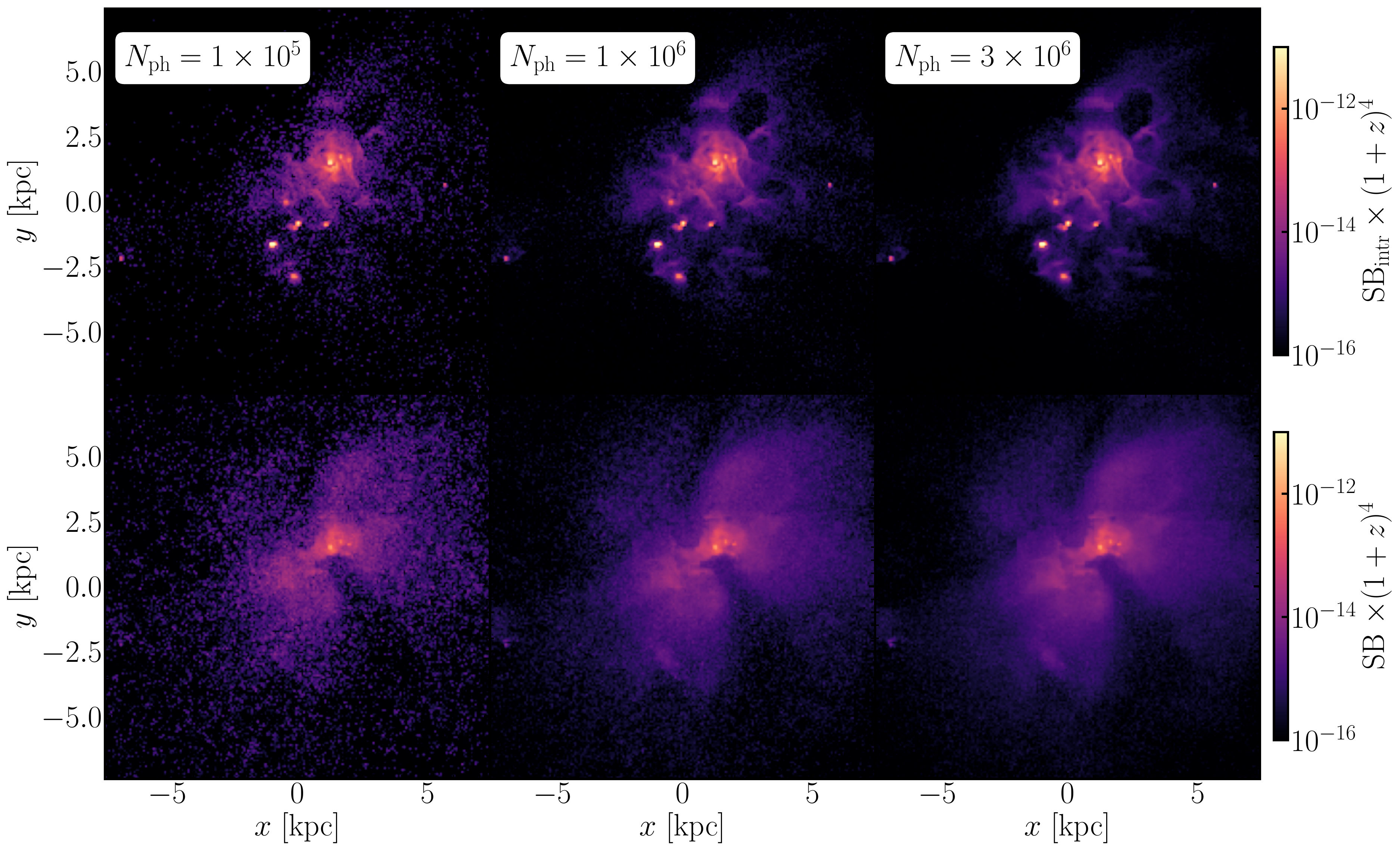}
    \caption{Convergence test of photon number counts for \lya images. The top row displays the intrinsic \lya images, while the bottom row presents the intrinsic scattered \lya images. From left to right, results are shown for photon number counts of $N_{\rm ph} = 1 \times 10^5$, $1 \times 10^6$, and $3 \times 10^6$, respectively.}
    \label{fig:img_conv_test}
\end{figure*}

\begin{figure}
    \centering
    \includegraphics[width=1.\linewidth]{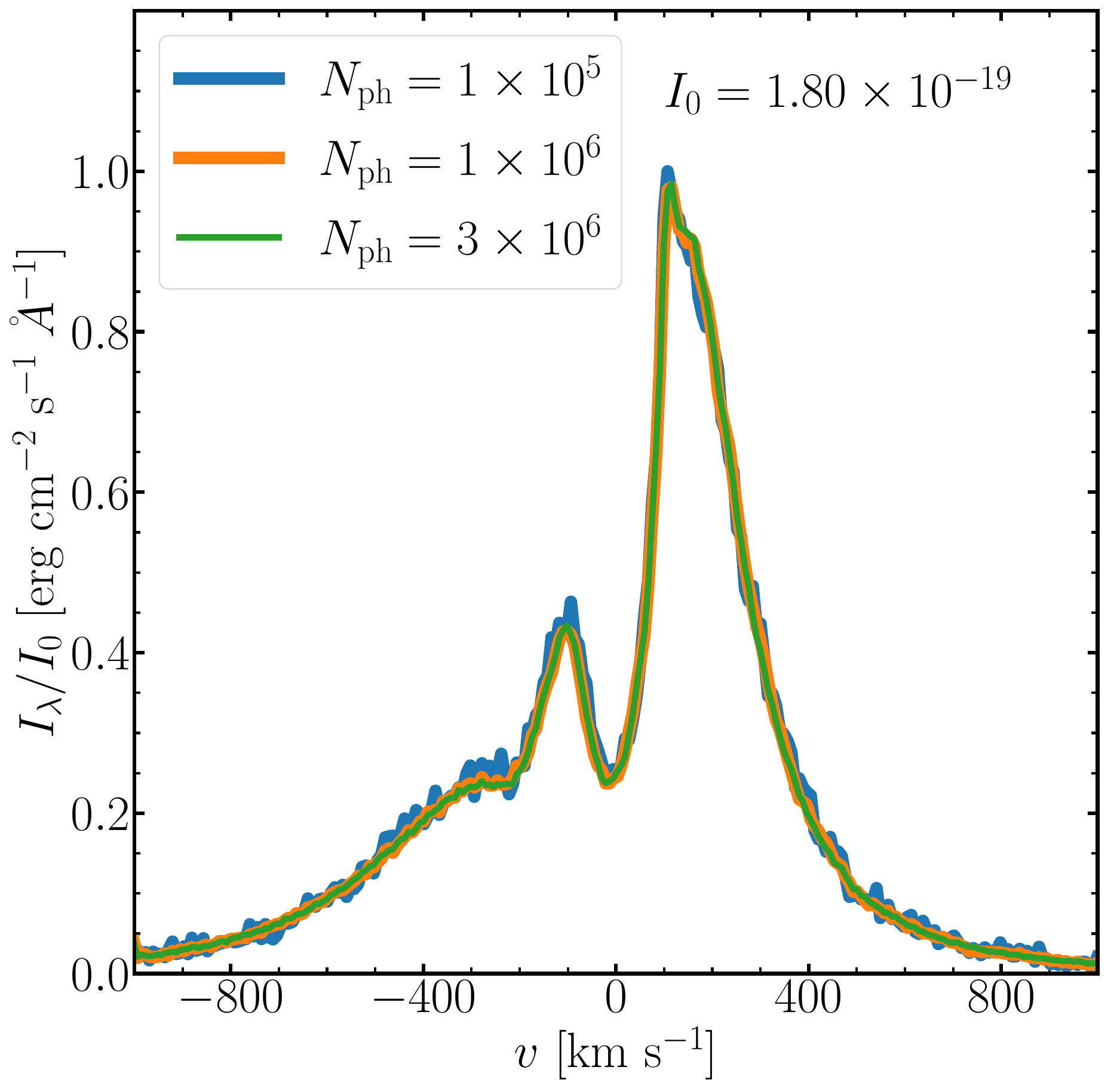}
    \caption{Convergence test of photon number counts for intrinsic and scattered \lya spectra. Results are shown for photon number counts of $N_{\rm ph} = 1 \times 10^5$, $1 \times 10^6$, and $3 \times 10^6$, represented by the blue, orange, and green curves, respectively. The spectra are normalized by $I_0$, with its value indicated in the top right corner.}
    \label{fig:spec_conv_test}
\end{figure}

MCRT codes such as \textsc{Rascas} track the trajectories of photon packets at spatial scales much smaller than the cell sizes used in hydrodynamic simulations (see, e.g., Section 3 of \citealt{Michel-Dansac20} for a detailed discussion). As a result, at fixed hydrodynamic resolution, MCRT simulations using sufficiently large numbers of photon packets can sample scattering events within galaxies thoroughly and produce convergent results. To assess the convergence of our \textsc{Rascas} MCRT simulations, we vary the number of photon packets from  $N_{\rm ph} = 10^5$, to $10^6$, and $3 \times 10^6$ in the \lya images shown in \autoref{fig:img_conv_test}. We find that all simulations converge in the central regions of the images, while the structures in the outskirts are well captured at resolutions of $N_{\rm ph} = 10^6$ (our fiducial value) and $3 \times 10^6$. A similar analysis is performed for the \lya spectra in \autoref{fig:spec_conv_test}. At our fiducial resolution, the MCRT simulations produce smooth \lya spectra with features that closely match those at the highest resolution. We therefore conclude that the images from our MCRT simulations achieve convergence at the fiducial photon packet number.


\bsp	
\label{lastpage}
\end{document}